\documentclass[fleqn,10pt,lineno]{wlpeerj}

\title{Self-Supervised Learning Methods and Applications in Medical Imaging Analysis: A Survey}

\author[1]{Saeed Shurrab}
\author[1]{Rehab Duwairi}
\affil[1]{Department of Computer Information Systems, Jordan University of Science and Technology, Irbid, Jordan}

\corrauthor[1]{Rehab Duwairi}{rehab@just.edu.jo}
\nolinenumbers
\begin{abstract}
\nolinenumbers
The scarcity of high-quality annotated medical imaging datasets is a major problem that collides with machine learning applications in the field of medical imaging analysis and impedes its advancement. Self-supervised learning is a recent training paradigm that enables learning robust representations without the need for human annotation which can be considered an effective solution for the scarcity of annotated medical data. This article reviews the state-of-the-art research directions in self-supervised learning approaches for image data with a concentration on their applications in the field of medical imaging analysis. The article covers a set of the most recent self-supervised learning methods from the computer vision field as they are applicable to the medical imaging analysis and categorize them as predictive, generative, and contrastive approaches. Moreover, the article covers 40 of the most recent research papers in the field of self-supervised learning in medical imaging analysis aiming at shedding the light on the recent innovation in the field. Finally, the article concludes with possible future research directions in the field.
\end{abstract}

\begin{document}

\flushbottom
\maketitle
\thispagestyle{empty}
\nolinenumbers
\section*{Introduction}

Medical image analysis is mainly concerned with processing and analyzing medical images, from different modalities, to extract useful information that help in making precise diagnosis \citep{anwar2018medical}. Medical images' analysis falls into four main tasks which were emerged from computer vision tasks and tailored for the medical filed.  These four tasks are classification, detection and localization, segmentation and registration \citep{altaf2019going}. Each of the mentioned tasks has its own methods and algorithms that help in understating and extracting useful information from the medical images.

The recent advancements in the artificial intelligence (AI) field brought significant improvements into the medical image analysis field by transforming it from a heuristic-based into a learning-based approach \citep{ker2017deep}. To elaborate more, learning-based analysis approaches aim at extracting useful information (features) that represent the input images in a way that fits the target medical image analysis task. In addition, features extraction can be accomplished manually (engineered) or automatically (learned) from the data \citep{sarhan2020machine}. While manual feature extraction is the main concern of the Statistical Machine Learning field, the Deep Learning field is mainly concerned with the automatic extraction of features and it is highly preferred.

A Convolutional Neural Network (CNN) is an example of deep learning models which deals with grid-based data such as images to learn the latent features in a hierarchical fashion from the fine level (lines and edges) to the complex level (objects). Mainly, seven types of layers constitute the structure of CNN, namely, convolutional layer, pooling layer, activation layer, fully connected layer, upsampling layer, dropout layer and batch normalization layer \citep{yamashita2018convolutional}. While both convolutional and pooling layers are responsible for features' extraction and aggregation, activation layer is responsible for non-linear transformation. The fully connected layer is responsible for mapping the learned features into an output vector of a certain dimension in case of classification tasks. In other cases, such as dense prediction, a transposed convolutional block is employed by the CNN which acts as upsampling layer, which is responsible for mapping the learned features into an output array of certain dimension \citep{5539957}. Lastly, both dropout and batch normalization layers are responsible for regularization. The process of optimizing the learnable layers in CNNs is accomplished through the gradient descent algorithm and its variants which aim at minimizing the difference between the network's output and the ground truth labels (i.e. minimize a loss function).

CNNs have become a popular choice in the field of medical image analysis and provided a tremendous progression into the various medical image analysis tasks due to their ability to deal with images in their raw formats; and the performance they provide which can be compared to the human performance at faster rates. Yet, CNNs are known to have an enormous number of trainable parameters to be estimated, usually in millions, to capture the underlying distribution in the input data. As a result, a relatively large amount of annotated data is required to achieve a better estimation of these parameters and enable performing supervised training \citep{MITCHELL202141}. 

Despite the remarkable success that CNNs have achieved in the medical image analysis field, there are some obstacles that hamper their advancement. Initially, building a large enough annotated medical dataset of high quality is expensive and time-consuming. In addition, unlike the natural scene image data which may be annotated by less skilled personnel, medical datasets require expert personnel with domain knowledge to accomplish the annotation process. Moreover, the annotation process is prone to patients’ privacy issues especially when working with specific disorders \citep{NEURIPS2020_d2dc6368}. Collectively, these factors render annotated data scarcity in terms of annotation and volume a major obstacle for machine learning applications in the medical field.

As an alternative solution, the concept of transfer learning came to the top of the table for situations where the amount of annotated data is relatively small. Transfer learning is the process of employing the knowledge that has been learned in a source task to another target task to improve the generalization and the performance \citep{goodfellow2016deep,torrey2010transfer}. The most common form of transfer learning, in the machine learning community, is built upon pre-trained state of the art models such as VGG \citep{DBLP:journals/corr/SimonyanZ14a}, GoogleNet \citep{szegedy2015going} , ResNet \citep{he2016deep} and DenseNet \citep{huang2017densely} which are trained on the giant labeled image datasets such as ImageNet \citep{deng2009imagenet}. ImageNet includes approximately 14 million natural images that belong to 22,000 visual categories and 1,000 labels \citep{krizhevsky2012imagenet}.

The employment of pre-trained models on ImageNet for medical applications is a controversial issue for three reasons. Firstly, the extracted features from the natural images domain may not be a good representation in the medical field due to the remarkable difference in features' distribution, resolution, and number of output labels between both domains. Secondly, ImageNet pre-trained models are over-parameterized models in terms of number of parameters when utilized for medical images analysis tasks. More clearly, ImageNet pre-trained models are designed to predict 1000 labels which makes them require a larger number of parameters, especially in the last layers to fit the 1000 classes. On the other hand, in the case of medical images, the number of classes may not exceed 10 classes, and hence, smaller models can be sufficient \citep{holmberg2020self,NEURIPS2019_eb1e7832}. Thirdly, ImageNet pre-trained models are primarily trained on 2D images while the vast majority of medical imaging modalities are 3D such as CT, MRI, and OCT. This renders the pre-trained models on the ImageNet dataset an infeasible solution. Despite that, a set of guidelines exists that mainly depends on the target dataset size and domain similarity when dealing with ImageNet pre-trained models for different domains \citep{karpathy2016cs231n}. Other approaches have been proposed to overcome such problems where Self-Supervised Learning is one of them.

Self-supervised learning is a recent learning paradigm that enables learning semantic features by generating supervisory signals from a pool of unlabeled data without the need for human annotation \citep{chen2019self}. The learned features from self-supervised learning are used for subsequent tasks where the amount of the annotated data is limited. From the unsupervised learning perspective, the self-supervised learning approach omits the need for manually annotated data, while the supervised perspective in the self-supervised learning approach is represented in model training with labels generated from the data itself \citep{liu2021self}.

\begin{figure}[ht]
    \centering
    \includegraphics[height=11cm]{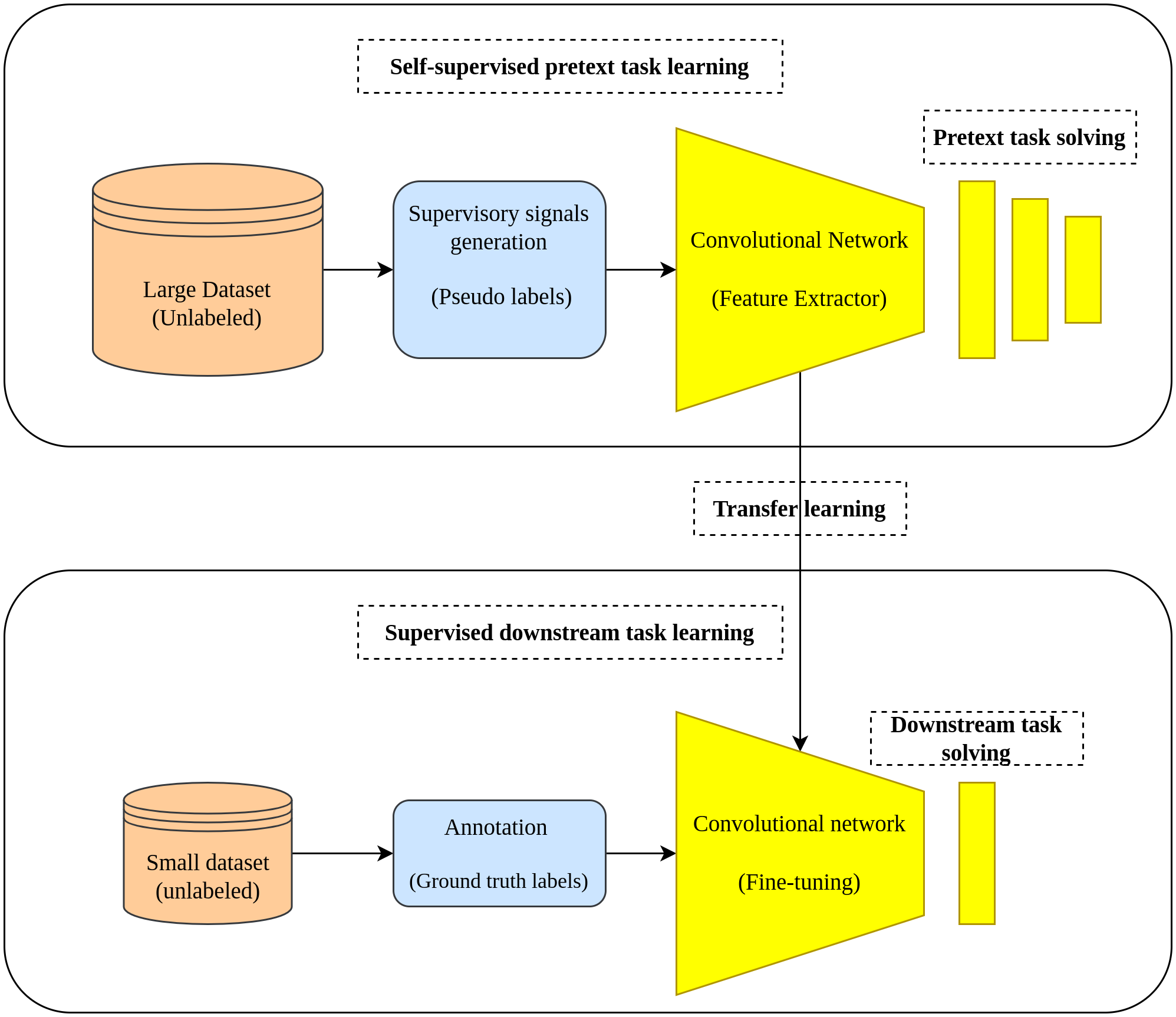}
    \caption[Self-supervised learning main workflow.]{Self-supervised learning main workflow. (top): Self-supervised learning scheme is applied by training an auxiliary task using pseudo labels generated from a large unlabeled dataset. (bottom): The learned representations are transferred from the pretext task to the down-stream task to accomplish the training on small amount of data with ground truth labels}
    \label{fig1}
\end{figure}

Two tasks characterize the learning pipeline in the self-supervised learning approach which are the pretext task and downstream task. In the pretext task where the self-supervised learning actually occurs, a model is learned in a supervised fashion using the unlabeled data by creating labels from the data in a way that enables the model to learn the useful representation from the data. In the downstream task, the learned representations from the pretext task is transferred as initial weights to the downstream task to accomplish its intended goal (fine-tuning) \citep{holmberg2020self}. Figure \ref{fig1} depicts the main workflow of the self-supervised learning approach.

Self-supervised learning became a popular choice in the field of medical image analysis - where the amount of the available annotated data is relatively small, while the unlabeled data is comparatively large. Several researches have demonstrated the effectiveness of the self-supervised learning approach throughout various medical image analysis tasks such as detection and classification \citep{lu2020@semi,li2021rotation,sriram2021covid}, detection and localization \citep{chen2019self, pmlr-v143-sowrirajan21a,nguyen2020self}, and segmentation tasks\citep{NEURIPS2020_d2dc6368,xie2020pgl,NEURIPS2020_949686ec}.

This paper aims at reviewing the state-of-the-art research directions in self-supervised learning approaches for image data with a concentration on their applications in medical images analysis. Annotated data scarcity is a major problem that hampers the advancement of machine learning applications in the medical field, and self-supervised learning can act as an effective solution for such a problem. Our main goal, in this paper, is to shed the light on the recent innovations in the field of self-supervised learning in medical imaging analysis by providing a high-quality overview of the recently developed methods in the field to enable the reader to be familiar with such approach which is interesting and quickly becoming the choice for several researchers in the machine/deep learning field. 

The prospective audience of this article includes, in the first place, machine/deep learning researchers and practitioners in medical images analysis and computer vision fields. Further, researchers and practitioners from the medical field who are interested in medical imaging analysis via machine learning approaches form a second group of the prospective audience. Lastly, any reader with an interest in machine learning applications, in general, is considered as the third group of the prospective audience. It is worthy to note that this survey was presented in a simplified manner to fit the various groups of the prospective audiences.   

Various research works, in the literature, have concentrated on self-supervised learning in computer vision per-se such as \citep{jing2020self, liu2021self,ohri2021review,jaiswal2021survey}, while other researches briefly reviewed the role of self-supervised learning in the analysis of medical images as part of deep learning applications in medical image analysis such as \citep{tajbakhsh2020embracing, chen2021recent}. To the best of our knowledge, this is the first survey on self-supervised learning applications in the field of medical images that aims at bridging the gap between computer vision and medical imaging fields. The key contributions of this paper can be summarized as follows: 
\begin{itemize}
    \item We provided a high-level overview of the state-of-the-art self-supervised learning methods in the computer vision field as they are general-purpose methods that can be used in the medical context. Further, we categorized these methods as predictive, generative, and contrastive self-supervised methods. 

    \item We covered and provided a high-level overview for a list of the 40 most recent and impactful research works in the field of self-supervised learning in medical imaging analysis. In addition, we categorized these works in the same way we categorized the computer vision tasks. Further, we included an additional category called multiple-tasks/multi-tasking to fit those researches that utilized multiple tasks simultaneously.
    \item We developed a GitHub repository\footnote{https://github.com/SaeedShurrab/awesome-self-supervised-learning-in-medical-imaging} called Awesome Self-Supervised Learning in Medical Imaging that would serve as a resource for the literature in the field which will be updated continuously.
\end{itemize}

The rest of this survey is organized as follows: the second section summarizes the literature selection methodology.
The third section provides an in-depth overview of the self-supervised learning approach and its methods. The fourth section reviews the recent self-supervised learning methods in medical imaging analysis. The fifth section compares the performance of the discussed self-supervised learning in medical imaging. The sixth section highlights some open challenges and the possible future research directions in the field, while the last section concludes the paper. Lastly, Appendix A lists the available implementation codes of the discussed research throughout this paper.

\section*{Survey Methodology}

This section summarizes the methodology followed, by the authors, to search for relevant literature on self-supervised learning applications in medical imaging analysis. This methodology includes the determination of literature sources, search keywords, inclusion/exclusion criteria, and papers selection criteria.    

\subsection*{Sources and keywords}

The first step in our methodology is to select the main sources of literature that will be used. As a result, we considered three bibliographic databases as primary sources of literature, namely:

\begin{itemize}
    \item IEEE Explore\footnote{http://ieeexplore.ieee.org/}
    \item ScienceDirect\footnote{https://www.sciencedirect.com/}
    \item Springer Link\footnote{http://link.springer.com/}
\end{itemize}

We focused our literature search on these resources as they include reputable journals and conferences that are mainly concerned with machine learning applications in medical imaging. On the other hand, we considered two additional sources of literature as secondary sources which are:

\begin{itemize}
    \item ArXiv Preprints\footnote{https://arxiv.org/}.
    \item The related works sections in the selected papers.
\end{itemize}

For searching keywords, we opted the terms \textit{self-supervised learning in medical imaging}, \textit{pretext tasks in medical imaging}, \textit{representation learning in medical imaging} and \textit{contrastive learning in medical imaging} to investigate the selected resources.

\subsection*{Inclusion/exclusion criteria}

Initially, we explored the literature in the field of self-supervised learning in medical image computing over the period 2017-2021, as this is the period where self-supervised learning started to creep into medical imaging analysis, with a high emphasis on the research works from the period 2019-2021 and excluded any other works outside this period. Further, we examined the titles and abstracts of the research articles resulting from querying the selected resources to judge the relevance of search results. As a result, we considered only research works that either have adopted a self-supervised learning approach directly to solve medical imaging tasks or presented a novel self-supervised learning approach in medical imaging that has not been seen before to our knowledge and excluded any other works of less relevance to our target. For self-supervised learning approaches from the computer vision field, we first explored the selected self-supervised learning research in medical imaging analysis literature and selected those methods that have been frequently used in the medical field even if they are not within the predefined period. We further added some additional state-of-the-art methods that have not been explored directly in the medical context and excluded any other methods. In addition, we kept refining our search results by selecting research articles that are published in journals or conferences with an impact factor of 3 or greater and excluded any other works published in venues with less impact factor than our threshold. For ArXiv preprints, we considered only those works cited in the selected published papers and excluded any other works. We further examined the affiliation and the research portfolio of the authors of these preprints before including their works. We also considered research works from outside the selected sources gathered by exploring the related works sections of the selected papers that are directly relevant to our target.

\subsection*{Papers selection}

\begin{figure}[!ht]
    \centering
    \includegraphics[scale=0.4, height=5.5cm,width=14.5cm]{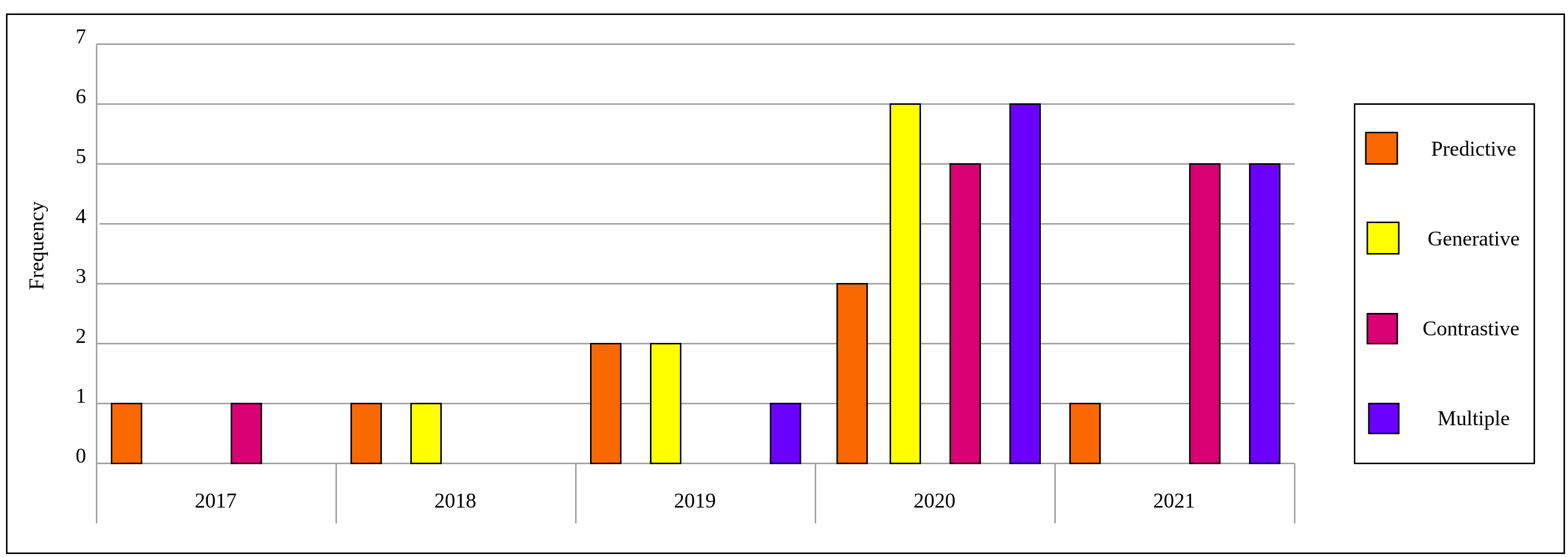}
    \caption{Distribution of selected publications by year and category for self-supervised learning in medical imaging.}
    \label{fig2}
\end{figure}

As a result of the predefined inclusion/exclusion criteria, we settled on 15 self-supervised learning approaches that have been developed on natural images and exploited in the medical context. For self-supervised learning in medical imaging,  we settled on 40 papers that relate directly to self-supervised learning applications in medical imaging analysis. Each of the selected papers has been reviewed thoroughly and a high-level overview is developed that focuses on the innovation in the self-supervised learning approach and presented throughout this survey. Figure \ref{fig2} depicts the distribution by year and category for the 40 papers in the field of self-supervised learning in medical imaging.

\section*{Self-supervised learning approaches}

The formulation of early self-supervised learning concepts appear in the work of \cite{bengio2007greedy} by training deep neural networks in an unsupervised greedy layer-wise fashion. The authors trained a single-layer auto-encoder for each layer one at a time (self-supervised learning). After training each layer in the network separately, the resulting weights of each layer are used as initial weights to train the whole network on the target task (fine-tuning). One of the prominent downsides of the greedy layer-wise training approach is the inability to secure a complete optimal solution by grouping sub-optimal ones \citep{goodfellow2016deep}. Further, the greedy layer-wise approach has been obsoleted by the emergence of end-to-end deep neural models that can be trained in a single run \citep{mao2020survey}. Despite that, the greedy layer-wise methodology formed the nucleus for what so-called nowadays self-supervised learning approach and opened the door for its applications in computer vision, natural language processing, robotics, and other fields. 

Pretext tasks play a central role in the self-supervised learning approach and act as its backbone. While the downstream task may differ according to the researchers' needs and targets, the pretext task can be common among different downstream tasks. For example, the same pretext task, e.g. convolutional auto-encoder, could be used to learn visual features for two different downstream tasks with different data. This property makes it helpful to categorize self-supervised learning approaches according to the nature of the pretext task. In this regard, we categorize self-supervised learning pretext tasks into three main categories including predictive, generative, and contrastive tasks. Such categorization aims at simplifying and grouping similar approaches together which in turn enables achieving a better understanding of the methods of each category. The upcoming sections introduce the reader to the most prominent methods for each category.


\subsection*{Predictive self-supervised learning}

The predictive self-supervised learning approach aims at learning robust representations from unlabeled data by treating the pretext task as a classification or regression problem. More clearly, each unlabeled image is assigned a pseudo label, these labels are generated from the data itself, which can be either categorical or numerical depending on the pretext task design specifications. As a trivial example, applying a certain transformation to the input image can be considered a pseudo label. Consequently, the role of the pretext task is to predict the this pseudo label correctly. It is worthy to note that pseudo labels must be carefully generated in order to enable learning robust representations from the data. Many predictive pretext tasks have been designed in the field of computer vision, the next sections illustrate in detail some of these methods.

\subsubsection*{Exemplar CNN}

\begin{figure}[!ht]
    \centering
    \includegraphics[scale=0.2, height=6.5cm]{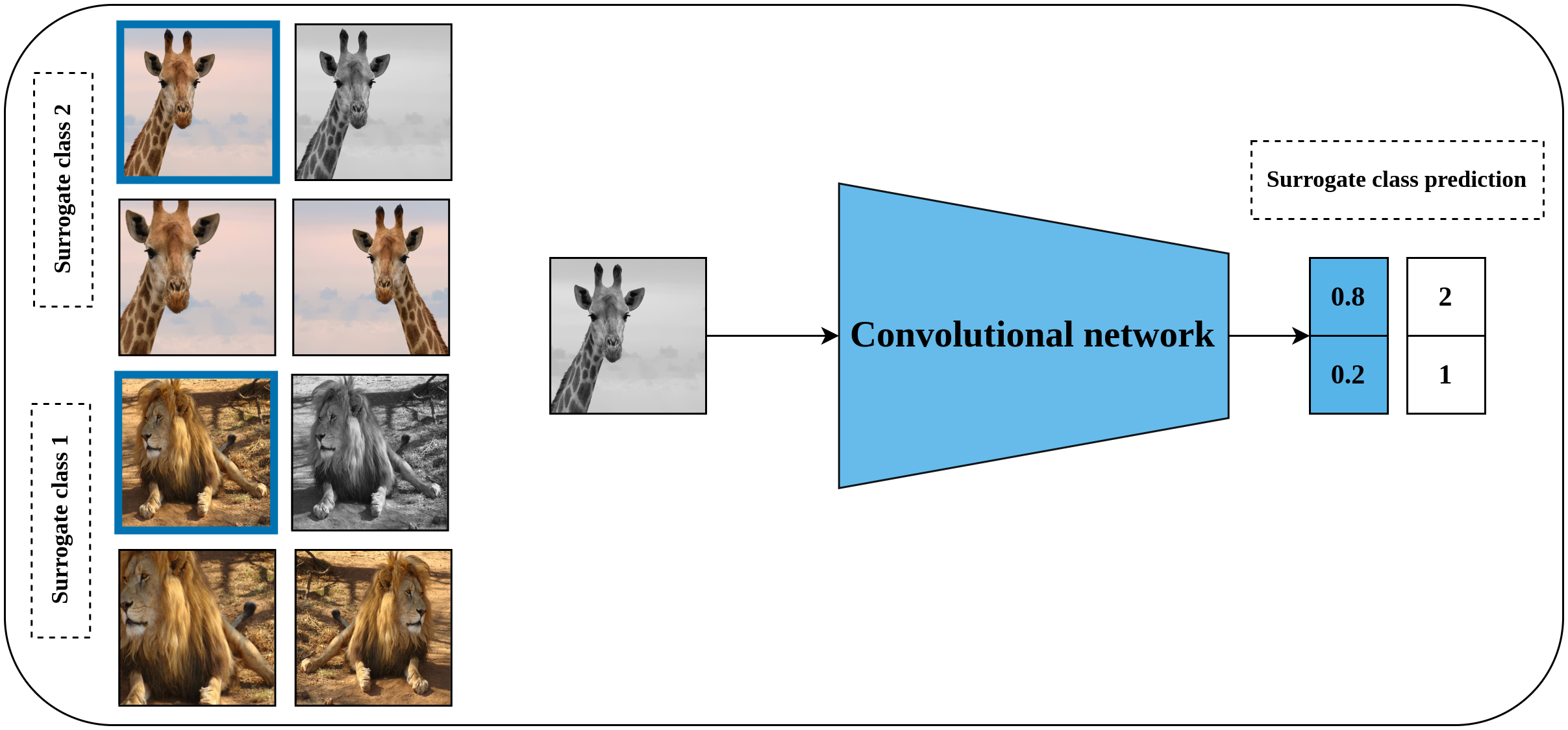}
    \caption{Illustration of the generation of surrogate classes for self-supervised features' learning with exemplar CNN. (left): The marked patch in blue represents exemplary patch cropped from a certain image in unlabeled dataset to serve as a seed for the surrogate class.  The remaining patches are a set of random augmentation operations applied to the seed patch to generate multiple images for the same surrogate class. 
    (right): A convolutional model is employed to learn representation by classifying the generated images into the specified surrogate classes. Image credit; upper: \href{https://www.pexels.com/photo/close-up-photography-of-giraffe-802112/}{Frans Van Heerden}, lower: \href{https://www.pexels.com/photo/brown-lion-730536/}{Gary Whyte}.}
    \label{fig3}
\end{figure}

Exemplar CNN is one of the earliest predictive self-supervised pretext models which was proposed by \cite{dosovitskiy2015discriminative}. Learning a good representation of the input data throughout the exemplar CNN method is hypothesized by the model's robustness to the applied transformations. To achieve this, a synthesized training dataset is created. This dataset consists of patches of objects or parts of the object with a size of $32\times32$ pixels which are cropped from the original images and they are called the exemplary patches. Following that, a set of predefined transformations including translation, scaling, rotation, contrast, and color adjustment are applied randomly to each generated patch as shown in Figure \ref{fig3}. Consequently, each seed patch along with its applied transformations forms a surrogate class in the training dataset. Following that, a convolutional neural network is trained to learn useful representations by learning to discriminate between the different surrogate classes in the synthesized dataset. 


\subsubsection*{Relative position prediction}
\begin{figure}[!ht]
    \centering
    \includegraphics[width=\textwidth, height=7cm]{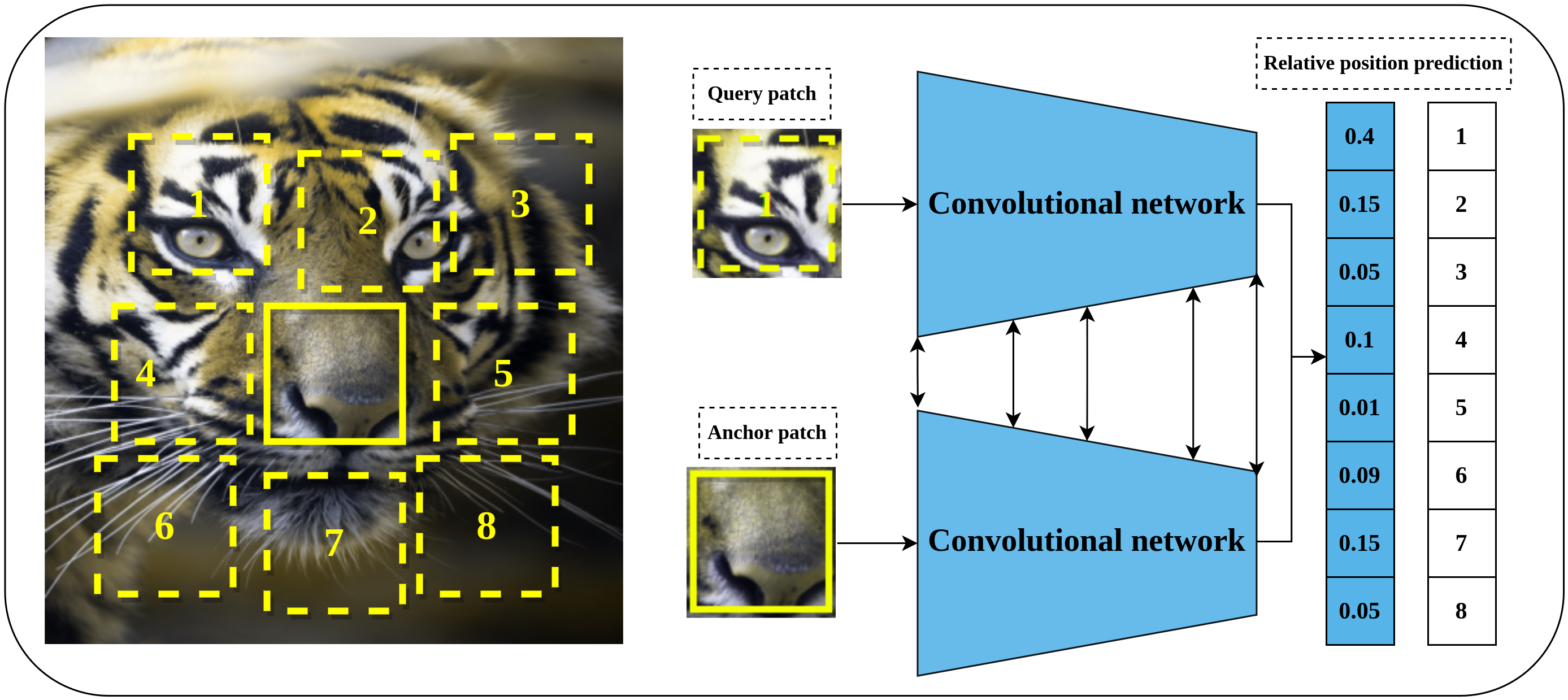}
    \caption{Illustration of self-supervised learning by relative position prediction task. (left): An image is divided into 9 patches where the central patch (the one without number) represents the anchor patch and the remaining 8 patches (delineated in dashed yellow lines) represent the query patches. (right): a training example is that consists of an anchor patch and query patch is passed to a late-fusion convolutional model which share weights between the two branches to predict the position of the query patch with respect to anchor patch. Image credit: \href{https://www.pexels.com/photo/fur-zoo-tiger-cat-7170713/}{Gabriele Brancati}.}
    \label{fig4}
\end{figure}

Relative position prediction is another predictive pretext task proposed by \cite{doersch2015unsupervised} that is inspired by the word embedding Skip-Gram model \citep{mikolov2013distributed} in natural language processing field. The main hypothesis of learning representations by relative position prediction is to understand the spatial context of the objects in the input image. The implementation details include dividing the input image into a $3\times3$ grid of patches as shown in Figure \ref{fig4}. The central patch is considered an anchor patch, while the remaining 8 patches are considered query patches. To increase the complexity and reduce the chance of learning shortcuts such as texture continuity and boundary patterns, a set of solutions was introduced including the addition of gaps and jitters to the patches, color channel processing by shifting certain channels to the gray-scale or partial channel dropping to avoid the chromatic aberration effect. Consequently, a late-fusion convolutional model is trained on a randomly sampled pair of patches of the central patch and query patch to predict the relative position of query patches with respect to the central patch.

\subsubsection*{Jigsaw puzzle}
\begin{figure}[!ht]
    \centering
    \includegraphics[scale=0.7]{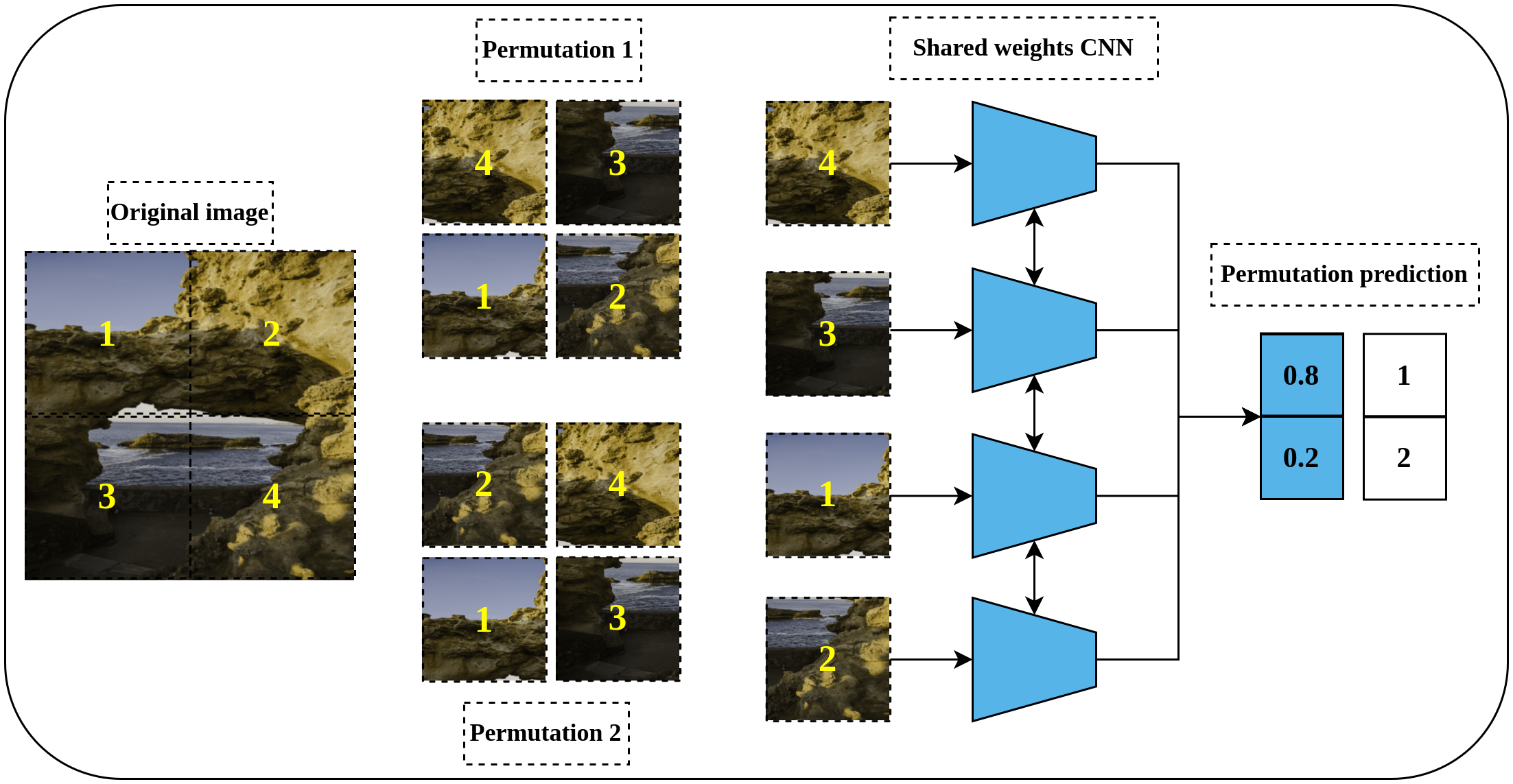}
    \caption{Illustration of a Jigsaw puzzle pretext task. (left): The Puzzle generation steps where an image is cropped into a set of patches that constitute the main blocks of the puzzle. The generated patches are shuffled according to a predefined set of permutations where each permutation has a specific index (permutation number). (right): a siamese network, with shared weights, takes the shuffled patches as input according to certain permutation and classifies them to the respective permutation index. Image credit: \href{https://www.pexels.com/photo/rocks-forming-arch-on-beach-11843536/}{Mathilde Langevin}.}
    \label{fig5}
\end{figure}

Solving a Jigsaw puzzle is another pretext task proposed by \cite{noroozi2016unsupervised} and inspired by the earlier work of \cite{doersch2015unsupervised} for relative position prediction. To solve a Jigsaw puzzle, a convolutional neural network is required to learn to restore a set of disordered patches, e.g. 9 patches, to their original spatial arrangement. For this purpose, a special convolutional network called Context-Free Networks (CFN) with siamese architecture and shared weights was proposed by the authors as shown in Figure \ref{fig5}. To train the network, a shuffled image with a random permutation of the 9 patches is fed to the network. But, for 9 patches there is $9! = 362,880$ possible permutations. To avoid such a large solution space, the authors limit the number of permutations to a predefined set of permutations with a certain index for each permutation. Lastly, the defined architecture's role is to produce a likelihood vector over the set of predefined indices that maximize the probability of the input permutation.

\subsubsection*{Rotation prediction}
\begin{figure}[!ht]
    \centering
    \includegraphics[width=\textwidth, height=7cm]{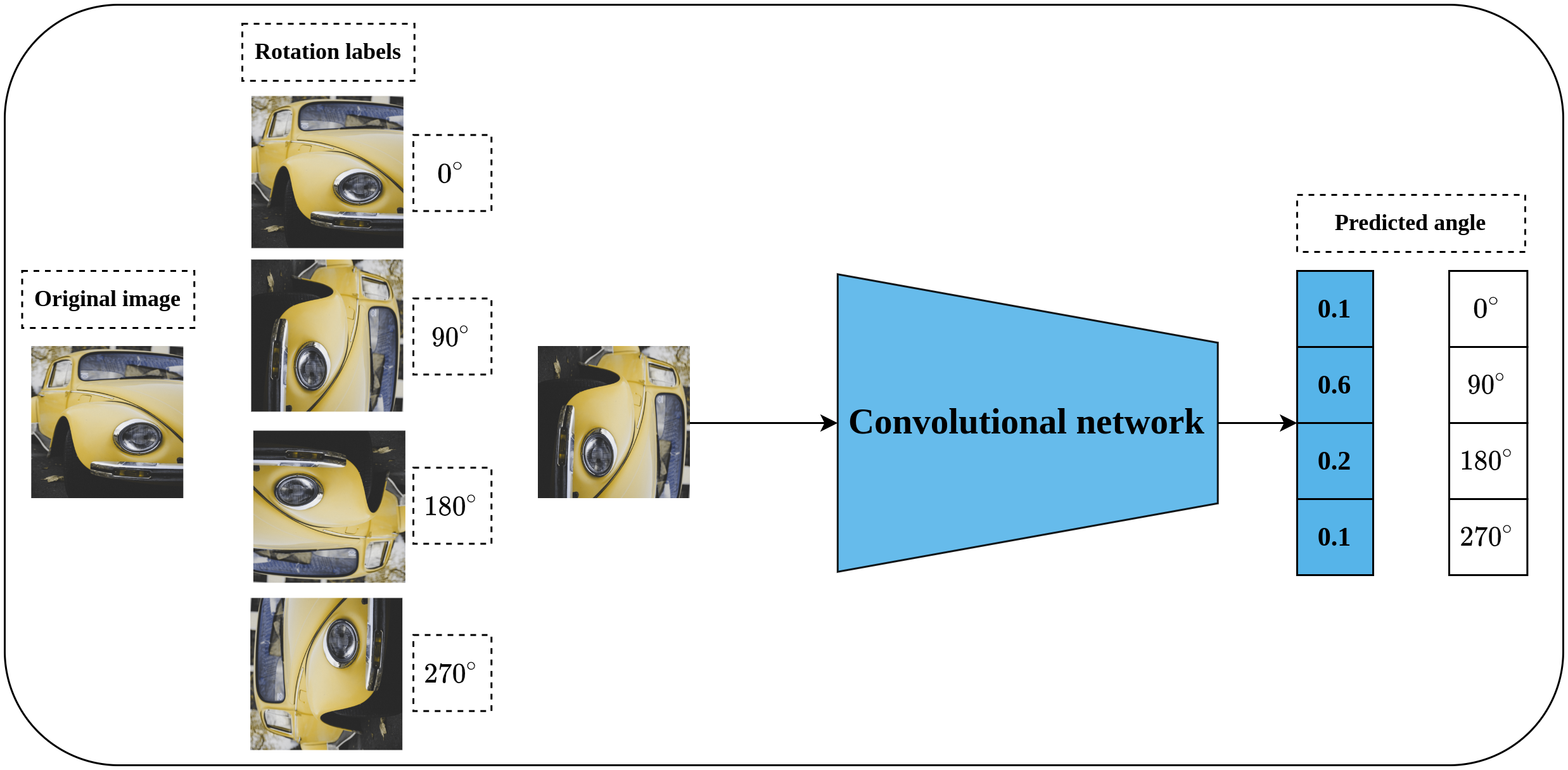}
    \caption{Illustration of rotation prediction pretext task. (left): Supervisory signals are generated from the data by applying a rotation angle in the range [$0^{\circ}$, $270^{\circ}$] with multiple of $90^{\circ}$ degrees to the input image. (right): The role of the network is to distinguish the applied rotation on the input image. Image credit: \href{https://www.pexels.com/photo/yellow-volkswagen-beetle-coupe-parked-on-gray-concrete-surface-1461804/}{Lilartsy}.}
    \label{fig6}
\end{figure}

Rotation prediction was first proposed by \citet{komodakis2018unsupervised} to learn visual representations in a self-supervised fashion. The main idea behind the rotation prediction task is to learn a convolutional model that can recognize the applied geometric transformation on the input image as shown in Figure \ref{fig6} in a simple classification problem. Geometric transformations are represented by applying rotation angles by multiple of $90^{\circ}$ to the input image which may fall into one of four categories including [$0^{\circ}$, $90^{\circ}$, $180^{\circ}$, $270^{\circ}$]. The main intuition behind the rotation prediction task is that enabling the convolutional network to learn to recognize the applied rotation to the input image is directly linked to the model's ability to learn the prominent objects in that image. To achieve this, the model needs to recognize the type and orientations of these objects in relation to the dominant geometric transformation to correctly learn the applied rotation. The same concepts hold for the human way of recognizing the rotation applied to a certain object in an image. For instance, to recognize a chair image which was rotated by $90^{\circ}$, a human needs to recognize the chair legs, base, back and their orientations. This way, rotation prediction enables learning semantic features by recognizing the orientations of images.


\subsection*{Generative self-supervised learning}

Generative self-supervised learning approach aims at learning the latent features in the input data by treating pretext tasks as a generative problem. The intuition behind generative pretext tasks is that the model can learn useful representations from unlabeled data by learning to regenerate the same input data or by learning to generate new examples from the same distribution of the input data. Generally, auto-encoder-based architectures generative adversarial networks are utilized in this category. Several generative pretext tasks have been proposed in the literature, the next sections illustrate, in detail, few of these methods.

\subsubsection*{Denoising auto-encoders}
\begin{figure}[!ht]
    \centering
    \includegraphics[width= 14cm, height=5cm]{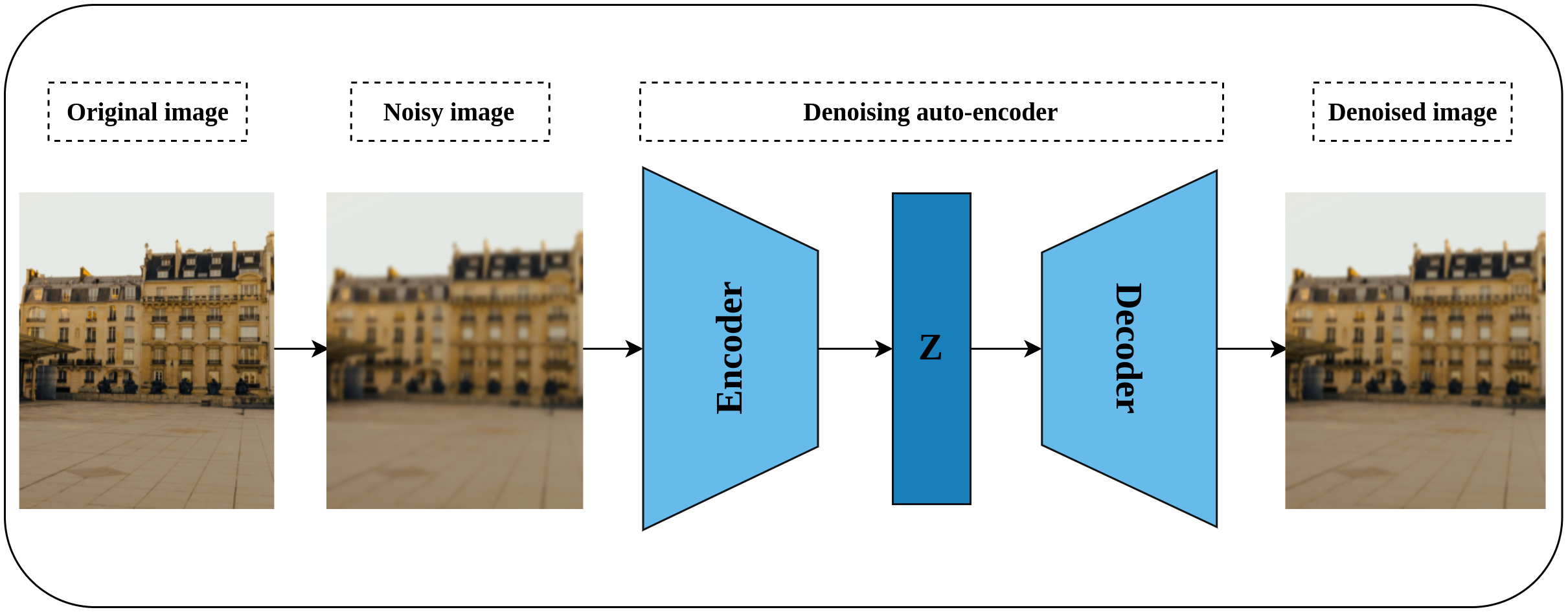}
    \caption{Illustration of Self-supervised features' learning using image denoising. (left): A noisy image is created by injecting noise to the original image. (middle) an auto-encoder model learns representations by compressing the noisy image into a latent space (Z) via the encoder network, while the decoder tries to reconstruct the compressed image from the latent space. (right): A denoised image close to the original image. Image credit: \href{https://www.pexels.com/photo/people-walking-on-sidewalk-near-brown-concrete-building-11387250/}{Céline}.}
    \label{fig7}
\end{figure}

Auto-encoders are special neural models whose main task is to reconstruct their input \citep{goodfellow2016deep}. The basic auto-encoder consists of two parts, namely, the encoder network and the decoder network. The encoder network plays the role of compressing the network's input into a latent dimensional space, while the decoder's role is to reconstruct the compressed input from the latent space \citep{tschannen2018recent}. After training the network, the decoder is discarded while the encoder is kept for further processing. Denoising auto-encoders are special models of auto-encoders proposed by \cite{vincent2008extracting} for representation learning through learning to reconstruct a noise-free output from noisy input. As shown in Figure \ref{fig7}, a noisy version of the original image is created by introducing certain types of noise including but not limited to Gaussian noise, Poisson noise, Uniform noise, and Impulsive noise. The noisy image is then passed to the auto-encoder to reconstruct the original image from the noisy image by minimizing the reconstruction loss. The intuition behind denoising auto-encoder is related to the human ability to correctly recognize the object type, in an image, even if a certain part of it is partially corrupted. This situation is true as long as the partial corruption does not affect the global view of the object. For a convolutional model, learning robust representations is linked to the model's ability to learn the semantic features that will enable restoring the original image from a noisy version.


\subsubsection*{Image inpainting}
\begin{figure}[!ht]
    \centering
    \includegraphics[width=\textwidth, height=4cm]{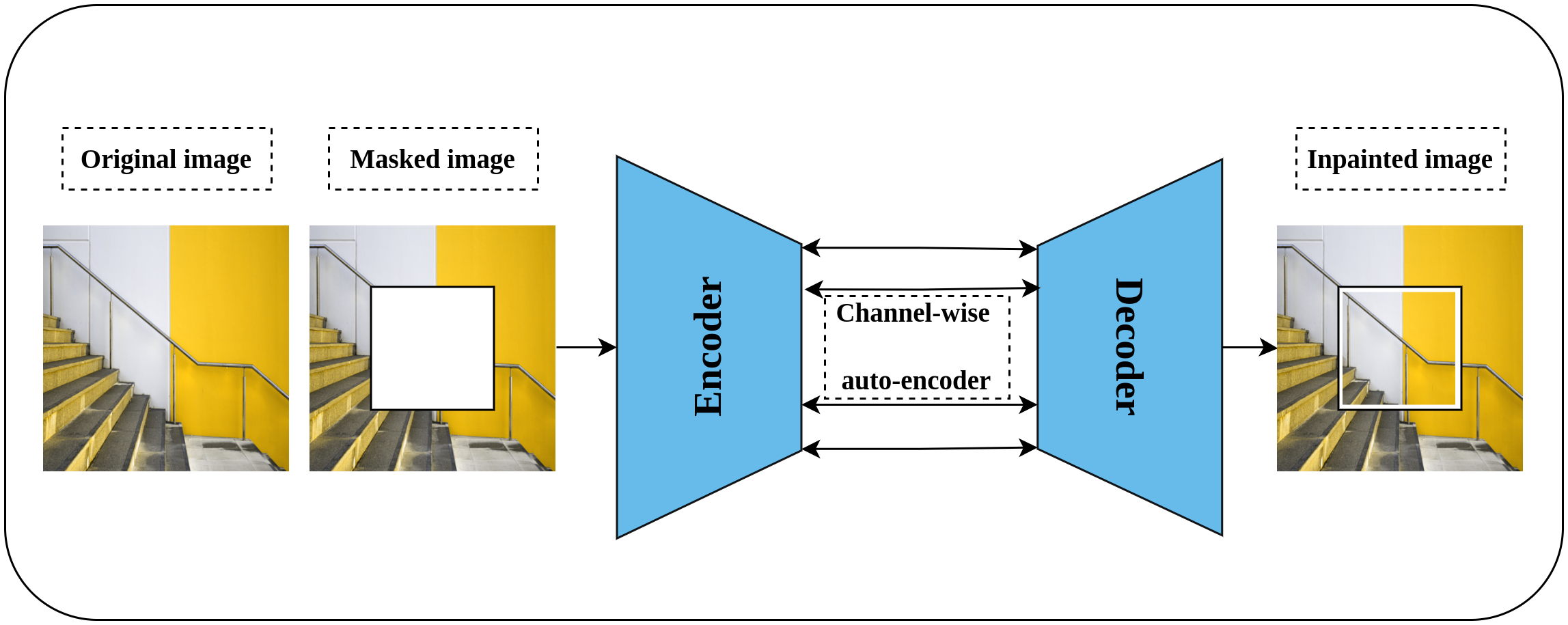}
    \caption{Illustration of context encoder model for Self-supervised features' learning. (left): An input image is modified by masking part of the image. (right): The context encoder learns useful representations by reconstructing the missing part in the masked image by minimizing the reconstruction and adversarial loss. Image credit: \href{https://www.pexels.com/photo/wood-stairs-light-water-6347788/}{Sam}}
    \label{fig8}
\end{figure}

Image inpainting or context encoder is a generative self-supervised pretext task proposed by \cite{pathak2016context} that aims to learn rich representations by fill-in-the-blank strategy. The intuition behind image inpainting is directly related to the human ability to complete the missing part of the image by observing the patterns in the surrounding pixels. Technically, part of the input image is cropped or masked, rather than introducing noise to it, and the role of the network is to complete the cropped part. Further, three forms of masking are proposed including central block, random blocks, and random region. An auto-encoder network and channel-wise fully connected latent space is employed for this task as shown in Figure \ref{fig8}. In addition, a combined loss function that integrates both reconstruction loss and adversarial loss \citep{goodfellow2014generative} is optimized throughout the training. The reconstruction loss, L2, is meant to hold the overall structure of the input image and the masked part, while the adversarial loss aims to improve the appearance of the predicted masked part. 


\subsubsection*{Image colorization}
\begin{figure}[!ht]
    \centering
    \includegraphics[width=\textwidth, height=5cm]{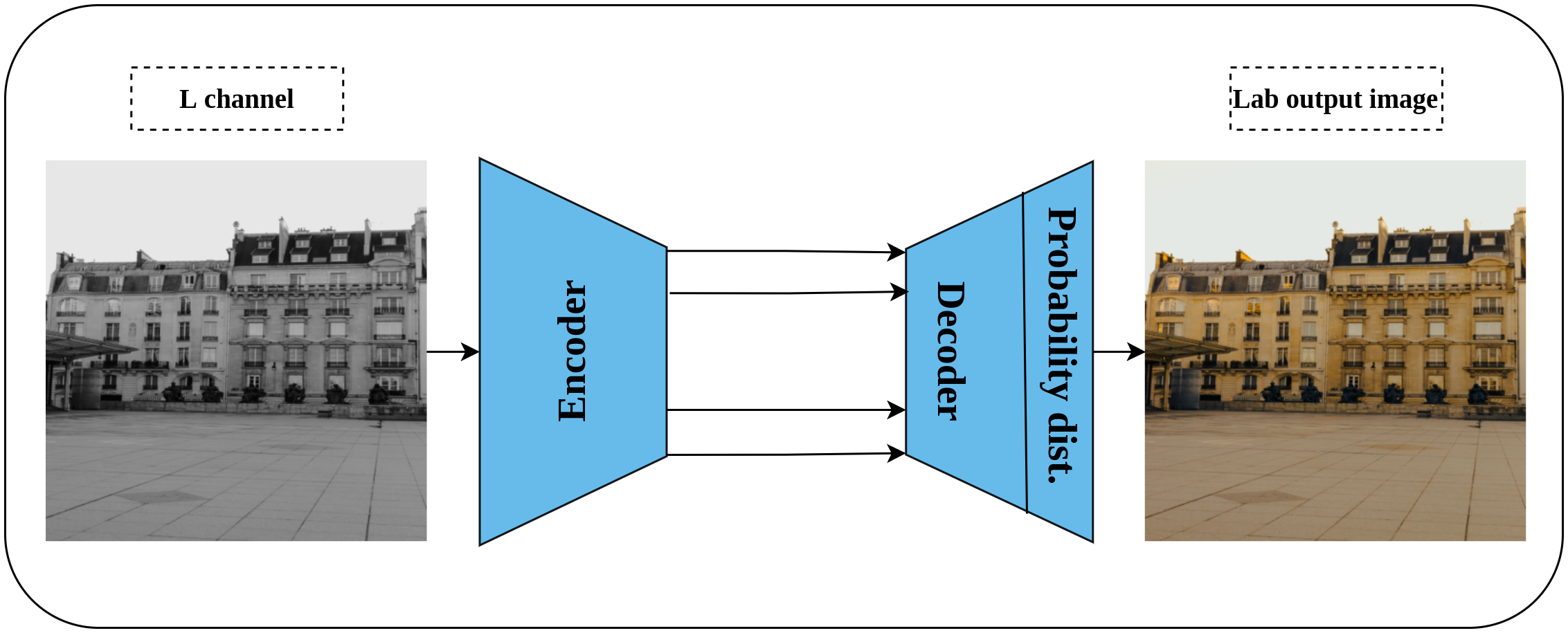}
    \caption{Illustration of image colorization pretext task.
    An encoder-decoder model is trained to predict the colored image from a gray scale image. The input is the L channel in Lab color space, while the channels a and b are used as supervisory signals. The last block indicates the color probability distribution for each pixel in the output image. Image credit: \href{https://www.pexels.com/photo/people-walking-on-sidewalk-near-brown-concrete-building-11387250/}{Céline}.}
    \label{fig9}
\end{figure}

Generation of a colorized image from a gray-scale one was proposed by \cite{zhang2016colorful} as a solution for automatic image colorization problem and self-supervised pretext task simultaneously. Lab color space is employed in this task rather than the RGB color space as it reflects the human color perception where the L channel represents the grayscale, while the a and b channels represent the color channels. Consequently, a convolution network is trained by taking the L channel as an input, and the channels a and b as supervisory signals - where the role of the network is to produce the input image in Lab color space as shown in Figure \ref{fig9}. Nonetheless, image colorization is multi-modal in nature which means that the same object may have different valid colors e.g. apple may be yellow, red, or green but not other colors. To compensate for this issue, the network is designed to predict the probability distribution of the possible colors for each pixel. In addition, a weighted cross-entropy Loss function is utilized to compensate for rare colors. Then, the annealed mean of the probability distribution is computed to produce the final colorization. The intuition behind the colorization task is that understanding the coloring scheme of the objects in the input images will result in learning rich representations about them.

\subsubsection*{Split-brain auto-encoder}

\begin{figure}[!ht]
    \centering
    \includegraphics[width=\textwidth, height=5cm]{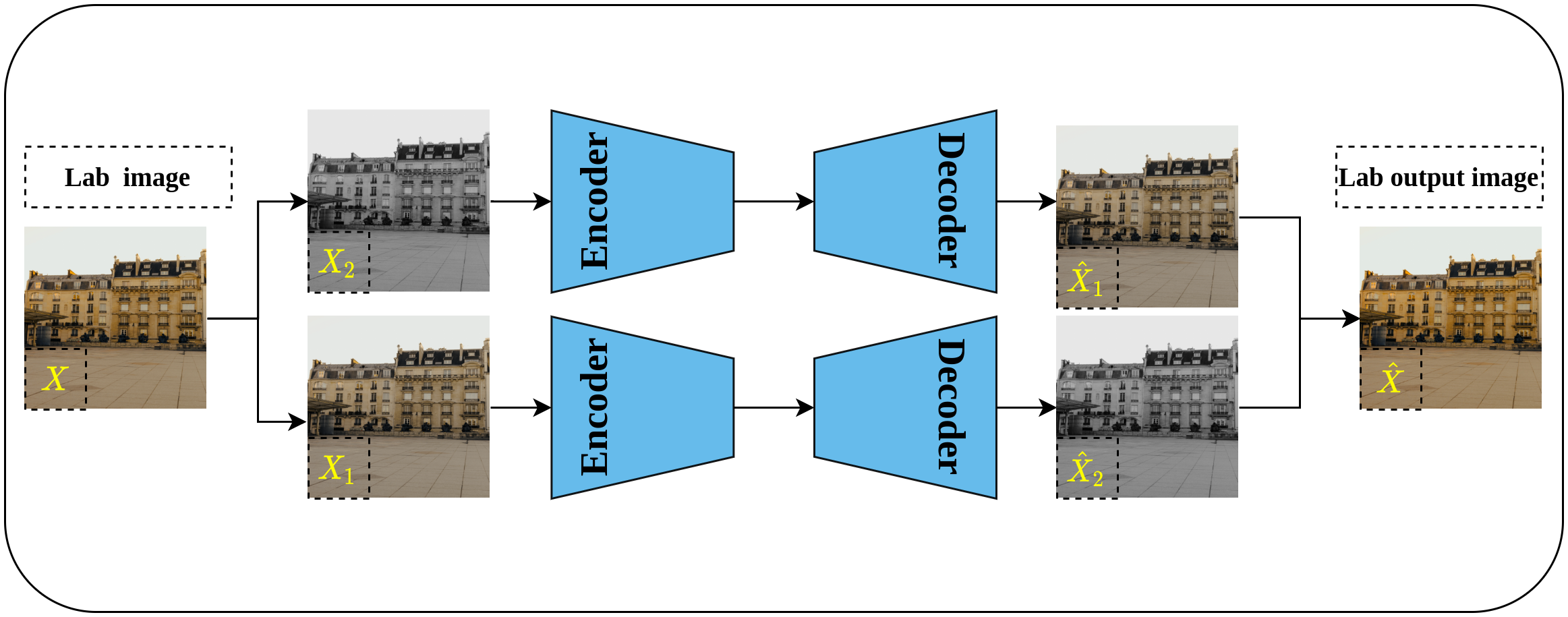}
    \caption{Illustration of split-brain auto-encoder pretext task. The input image $X$ is separated by channels as color channels  $X_1$ and gray-scale channel $X_2$. Two disjoint networks $F_1$ and $F_2$ are trained to predict the missing components in their inputs. $F_1$ predicts the gray-scale channel $\hat{X_2}$ from the color channels $X_1$, while $F_2$ predicts the color channels $\hat{X_1}$ from the gray-scale channels $X_2$. The outputs of both networks are grouped to produce the recolored image $\hat{X}$. Image credit: \href{https://www.pexels.com/photo/people-walking-on-sidewalk-near-brown-concrete-building-11387250/}{Céline}.}
    \label{fig10}
\end{figure}

Split-brain auto-encoder is another pretext task proposed by \cite{zhang2017split} and extended their earlier work on image colorization. The main idea behind the split-brain auto-encoder is to obtain useful representations by learning to generate a portion of the data from the remaining data. By translating this idea to the image data in Lab* color space, the gray-scale channel L can be generated from the color channels a and b and vice versa. This process is accomplished through modifying the traditional auto-encoder architecture by adding two splits to the network as shown in Figure \ref{fig10} - where each disjoint split learns the underlying representations from the input data as described previously. Eventually, the output of both splits is aggregated throughout concatenation to produce the final output of the network. The authors stated that learning from both gray-scale and color channels simultaneously rather than single-channel as in colorization problems would enable better representations learning. This is because the split-brain architecture is able to learn color-related information which is not the case in the colorization task which learns features only from gray-scale input.

\subsubsection*{Deep Convolutional GAN}
Generative adversarial networks (GAN) are a class of deep learning generative models that use random noisy input to generate new data which mimics the real training data. Typically, GAN architecture consists of two networks, namely, the generator network and the discriminator network. The role of the generator is to convert the random noisy input into an imitation of the real data while the role of the discriminator is to distinguish whether the generator output is real or fake. Both networks are trained in a competing way until the discriminator is not able to differentiate between real and fake images \citep{goodfellow2014generative}.

\begin{figure}[!ht]
    \centering
    \includegraphics[width=14cm, height=7cm]{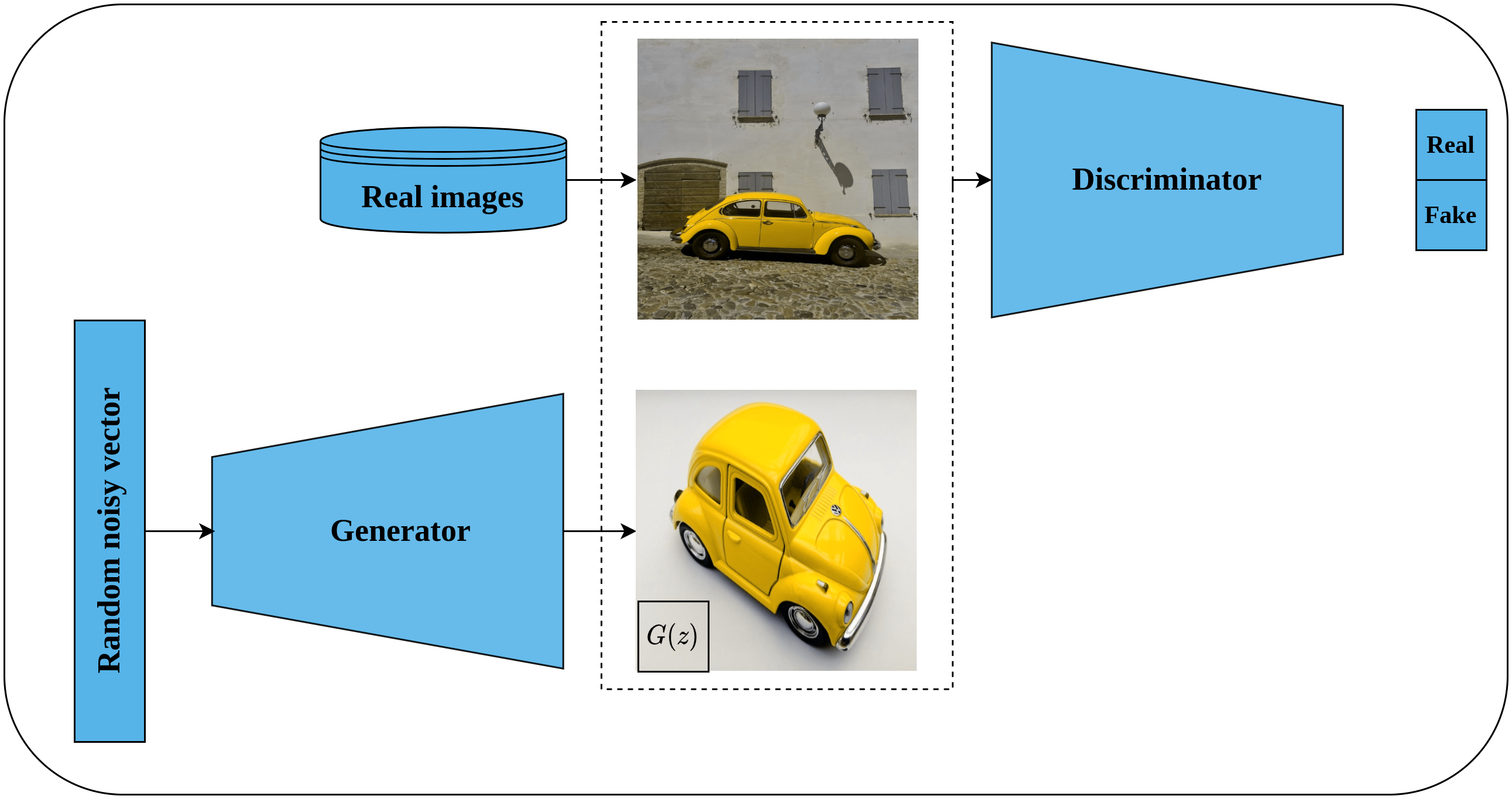}
    \caption{Illustration of deep convolutional GAN architecture. (left): A generator network tries to generate fake images using random noisy vector. (right): A discriminator network takes the generated images from generator network as well as real images from the same distribution and classifies them as real or fake until being not able to discriminate both sources. Image credit; upper: \href{https://www.pexels.com/photo/yellow-volkswagen-car-parked-on-the-street-10183798/}{Beatrice Gemmi}, lower: \href{https://www.pexels.com/photo/yellow-volkswagen-beetle-toy-car-on-white-background-3828253/}{Mike B}.}
    \label{fig11}
\end{figure}

Deep convolutional GAN, or DCGAN for short, is an extension of GAN proposed by \cite{DBLP:unsupjournals/corr/RadfordMC15} as an unsupervised representations learning architecture for image data. DCGAN is considered the first successful attempt to scale GAN with convolutional neural networks as opposed to the earlier work of \cite{goodfellow2014generative} which is based on multi-layer perceptron architecture. Further, the authors provided architectural guidelines for designing a stable DCGAN, such as replacing the pooling layer with a stridden convolutional layer for discriminator, and fractionally strided convolution for generator. Also, employing Batch normalization \citep{ioffe2015batch} in the generator and discriminator networks, removing  fully connected layers, using ReLU activation \citep{nair2010rectified} for all generator layers except the output layer which is Thanh activation. LeakyReLU activation \citep{maas2013rectifier} was recommended for all layers in the discriminator network. Figure \ref{fig11} depicts the generator network architecture as designed by the authors. The authors evaluated the quality of the learned features by DCGAN discriminator by performing an image classification task which showed superior performance in comparison to other unsupervised methods and opened the door for exploiting GAN-based models in pretext tasks.

\subsubsection*{Bi-directional GAN}
\begin{figure}[!ht]
    \centering
    \includegraphics[width=11cm, height=7cm]{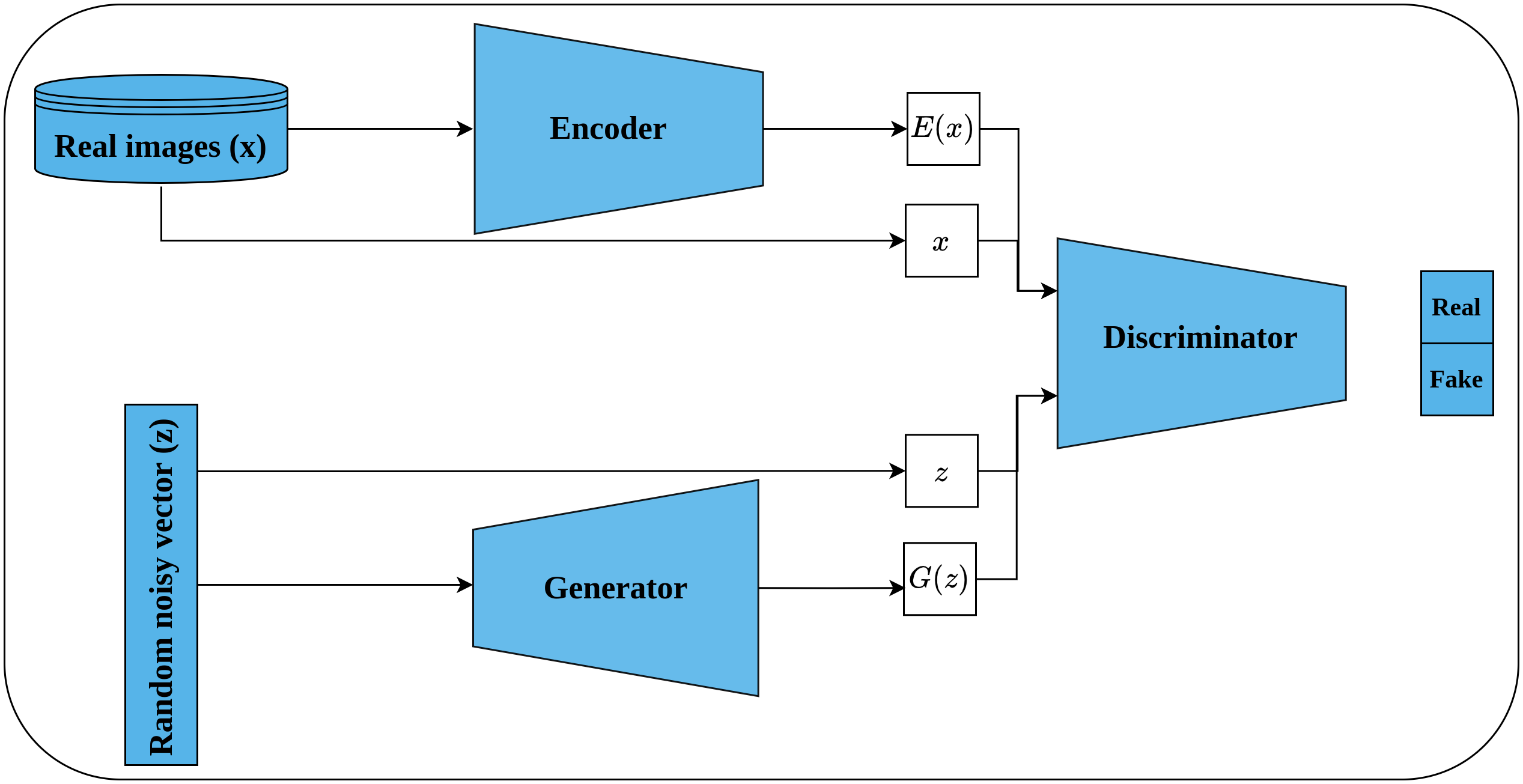}
    \caption{Illustration  of self-supervised features learning using Bi-directional GAN.(lower left): A generator network that generates a fake image $G(z)$ from random latent space $z$. (upper left): An encoder network that maps real image $x$ into a latent space $E(x)$. (right): The discriminator network takes as input a tuple of latent space and an image; and classify them as real or fake.}
    \label{fig12}
\end{figure}

Bi-directional GAN (BiGAN) is another generative unsupervised learning architecture proposed by \cite{donahue2016adversarial} that extended the earlier work of \cite{DBLP:unsupjournals/corr/RadfordMC15}. BiGAN introduces an encoder $E$ which maps an image  $x$ back to latent space $E(x)$ (called inverse mapping). The generator decodes random latent space $z$ to produce a fake image $G(z)$. Consequently, the discriminator $D$ takes, as an input, a tuple of latent space and an image which may be either $(G(z),z)$ or $(x,E(x))$ as shown in Figure \ref{fig12}. The role of the discriminator is to discriminate whether its input tuple is real or fake. The intuition behind incorporating the latent space along with the input image is to serve as free labels generated from the data without supervision in a similar way to learning representations by full supervision. The authors stated that both $E$ and $G$ are completely separated modules that do not communicate with each other during the training. Hence, both modules should learn to invert each other to be able to beat the discriminator. When training is complete, the learned representations, by the encoder, can be transferred to the downstream tasks.



\subsection*{Contrastive self-supervised learning}

Contrastive self-supervised learning is a recent representation learning approach that aims at developing robust representations from the input data by learning to differentiate between the similar (\emph{positive}) pairs and the dissimilar (\emph{negative}) pairs or by maximizing the agreement between a pair of positive views depending on the design specifications of the contrastive learning architecture. Positive examples are generated by applying a set of random augmentations to an input image which results in two transformed views of the same image, while negative examples are any other images different from the transformed views. The positive examples are assumed to be slightly different but preserve the global features of the input image which makes the similarity between them higher. Lastly, a contrastive model is trained to maximize the similarity between the positive pairs and minimize it with the negative pairs in case of using them. The next sections illustrate five contrastive learning approaches.

\subsubsection*{Contrastive predictive coding}

Contrastive predictive coding (CPC), is a contrastive unsupervised representations learning proposed by \cite{DBLP:cpcjournals/corr/abs-1807-03748} that can fit not only image data but also text, and audio. The main intuition behind CPC is to develop a compact representation that maximizes the mutual information between the context $C$ and the target $X$, rather than predicting the target $X$ directly from $C$ as it is the case with generative models. Such approach enables  learning rich representations as it ignores low-level information about the objects in the input data. The architecture of the CPC network consists of three components, namely, an encoder network which is responsible for converting the input into a compact latent variable $Z_t$; an auto-regressive network which is responsible for producing the context $C_t$ out of the encoded latent variables and generating future predictions; and the contrastive loss function, which is called InfoNCE which is formulated based on the Noise-Contrastive Estimation loss function (NCE) \citep{gutmann2010noise}.

\begin{figure}[!ht]
    \centering
    \includegraphics[width=\textwidth, height=5cm]{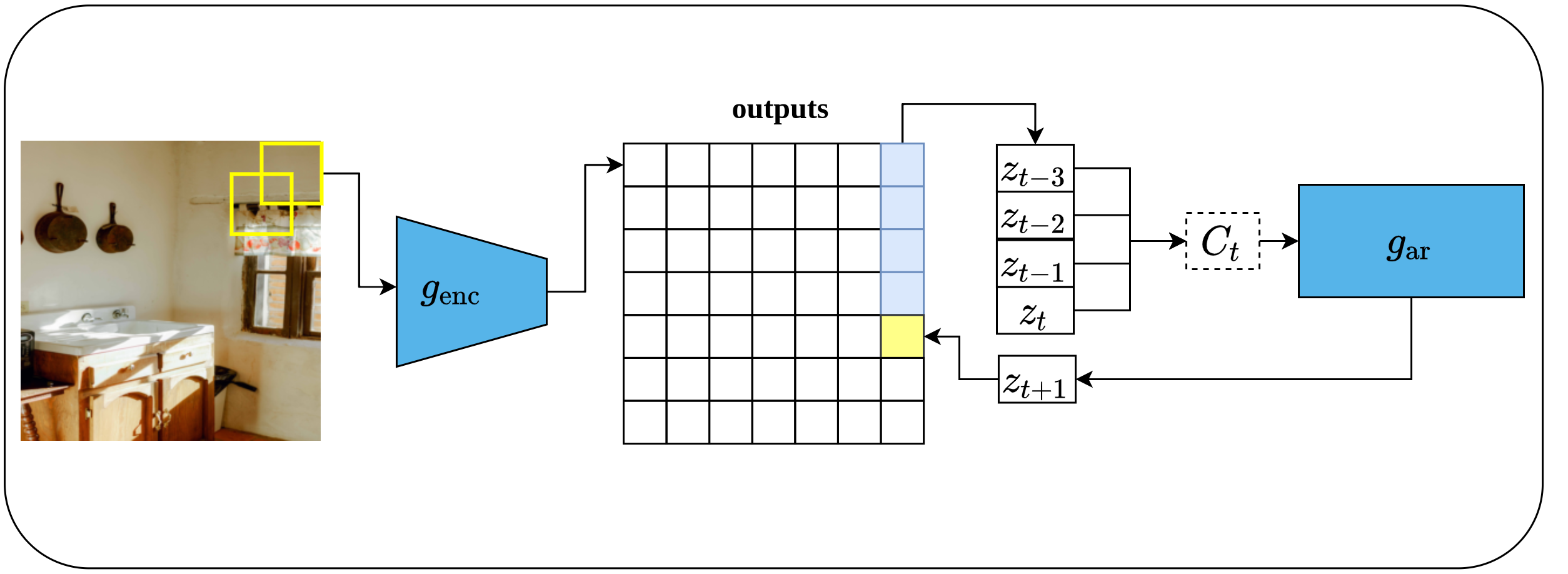}
    \caption{Illustration of features learning using contrastive predicting coding applied on image data. (left): The input image is rearranged into a grid of overlapping patches of size $7\times7$. Each crop is then encoded via a convolutional network $g_{\mathrm{enc}}$ (right): An auto-regressive model is used to make the predictions in top-to-bottom fashion. Image credit: \href{https://www.pexels.com/photo/photo-of-vintage-kitchen-11483728/}{Ali Alcántara}.}
    \label{fig13}
\end{figure}

To apply CPC on visual data (images), an input image of size $256\times256$ pixels is cropped into patches of size 64x64 pixels with an overlap of 32 pixels with respect to the height and width. This results in a grid of patches of size $7\times7$. Consequently, each patch is encoded via ResNet-101-v2 \citep{he2016identity} encoder into a vector $Z_{t}$ of size 1024 while the whole image forms 7x7x1024 tensor as shown in Figure \ref{fig13}. Following that, a PixelCNN architecture \citep{10.5555/3157382.3157633} is employed as an auto-regressor that generates a context vector $C_t\  \forall\  Z_{\leq t}$ which generate future predictions $Z_{t+k}$ in a top-down fashion and in a way that maximizes the mutual information between the context and predictions. Lastly, the InfoNCE loss function contrasts between the predicted patch and all other negative patches; these may come from other locations in the input image or other images in the same mini-batch. CPC V2 \cite{henaff2020data} is the second version of CPC which comes with several enhancements over the original CPC.

\subsubsection*{Momentum contrast}
\begin{figure}[!ht]
    \centering
    \includegraphics[width=12cm, height=7cm]{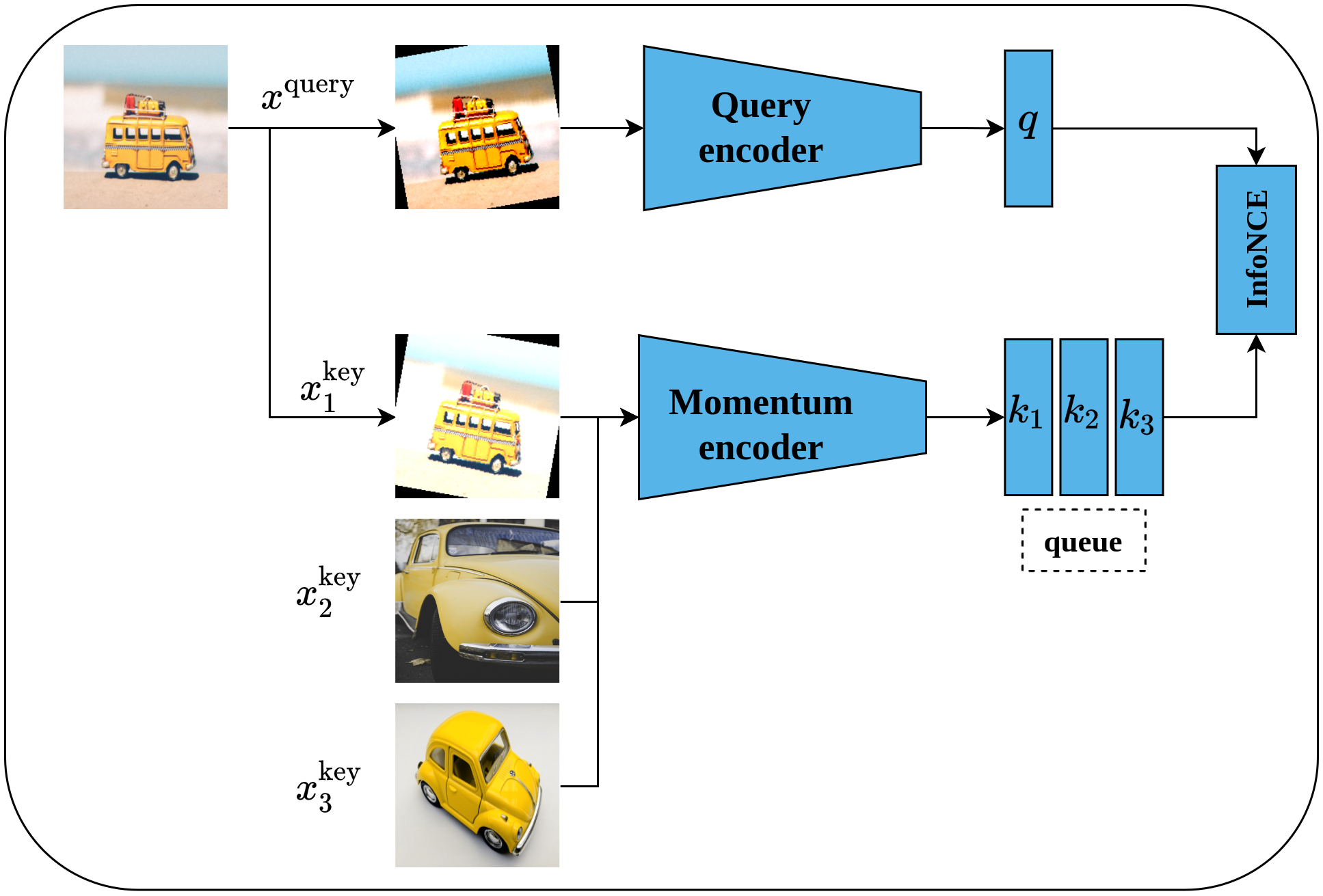}
    \caption{Illustration of momentum contrast framework. Image credit; upper: \href{https://www.pexels.com/photo/yellow-van-die-cast-386025/}{Nubia Navarro}, middle: \href{https://www.pexels.com/photo/yellow-volkswagen-beetle-coupe-parked-on-gray-concrete-surface-1461804/}{Lilartsy}, lower: \href{https://www.pexels.com/photo/yellow-volkswagen-beetle-toy-car-on-white-background-3828253/}{Mike B}.}
    \label{fig14}
\end{figure}

Momentum contrast (MoCo) is another self-supervised contrastive learning approach proposed by \cite{he2020momentum}. MoCo framework is inspired by \textit{dynamic dictionary-lookup} and \textit{queues} ideas. The main intuition behind MoCo is to perform a lookup operation using query image encoding in a dictionary that contains keys represented as images' encodings. Learning robust representations is enabled by learning to maximize the similarity between the encoding of the query image and the encoding of its matching key; and to minimize the similarity between the encoding of the query image and non-matching keys. Technically, MoCo architecture consists of two networks, namely, query-encoder and momentum-encoder as shown in Figure \ref{fig14}. The query-encoder role is to generate a features vector $q$ from the query image $x^\mathrm{query}$. The momentum-encoder  acts as a dictionary of data samples (whole images or patches $x_i^\mathrm{key}$) generated form encodings $k_i$ of features' vectors. Moco maintains a dynamic dictionary which should be of large size and consistent. The dictionary is designed as a queue of feature vectors' encodings $k_i$, where the present mini-batch enters the queue while the outdated mini-batches leave the queue in a First-In-First-Out fashion. Moreover, the dictionary size is not restricted to the mini-batch size but can be larger. On the other side, as the keys of the dictionary are derived from a group of previous mini-batches, they need to be updated regularly to maintain the consistency property. A momentum update of keys based on values of parameters of the query-encoder is proposed by the authors - where only the query-encoder parameters are updated by back-propagation, while the momentum-encoder is updated consequently using moving average; this allows it to be updated slowly and in a smoother fashion than the query-encoder. MoCo network is trained by minimizing the InfoNCE contrastive loss function \citep{DBLP:cpcjournals/corr/abs-1807-03748}. 


\subsubsection*{Simple framework for contrastive learning of visual representations}
\begin{figure}[!ht]
    \centering
    \includegraphics[scale=0.2, height=7cm]{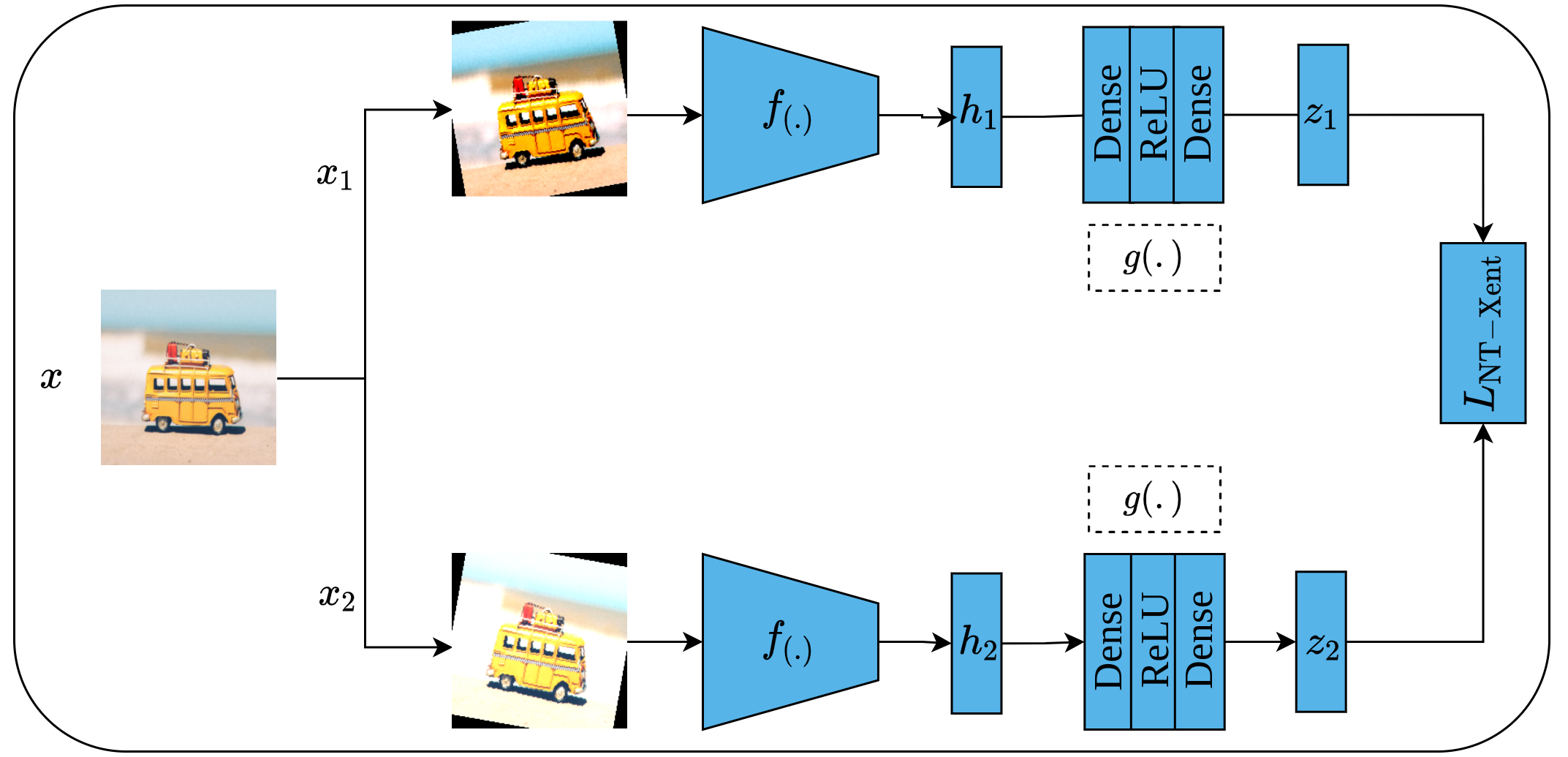}
    \caption{Self-supervised features learning by SimCLR. Image credit: \href{https://www.pexels.com/photo/yellow-van-die-cast-386025/}{Nubia Navarro}.}
    \label{fig15}
\end{figure}

Another contrastive learning approach is the simple framework for contrastive learning of visual representations, or SimCLR for short, which was proposed by \cite{chen2020simple}. As its name implies, SimCLR depends mainly on two simple ideas including heavy data augmentation techniques that result in correlated views for the same input, in addition to a large batch size that includes a large set of negative examples. Furthermore, SimCLR omits the need for additional functionalities as seen in CPC \citep{DBLP:cpcjournals/corr/abs-1807-03748} and MoCo \citep{he2020momentum}. To elaborate more on the SimCLR approach, a set of random transformations $\tau$ including cropping and resizing, flipping, rotation, color distortion, and Gaussian blur are applied to the input image $x$  which results in a pair of positively correlated views of the same image $(\tilde{x_i},\tilde{x_j})$ as shown in Figure \ref{fig15}. Consequently, both views are passed into a pair of convolutional encoders $f(.)$, ResNet50 \citep{he2016deep}, to learn representations of both views, which are denoted as $(h_i,h_j)$, respectively. Following that, the generated representations are passed to a pair of projection heads $g(.)$ which consist of two dense layers with ReLU activation \citep{nair2010rectified} for the first layer; and linear activation for the second layer. This results in a pair of feature vectors $(z_i,z_j)$. Lastly, the InfoNCE contrastive loss \citep{DBLP:cpcjournals/corr/abs-1807-03748,he2020momentum} named as Normalized Temperature-Scaled Cross-Entropy Loss (\textit{NT-Xent}) by the authors is employed to optimize the whole architecture based on the generated embedding $(z_i,z_j)$ by maximizing the agreement between the positive pairs of augmented images, while minimizing it for other images in the same batch (negative samples). when training is complete, the dense layers are discarded while the convolutional encoders are kept to be utilized in downstream tasks. 

\subsubsection*{Bootstrap your own latent}
\begin{figure}[!ht]
    \centering
    \includegraphics[width=14cm, height=7cm]{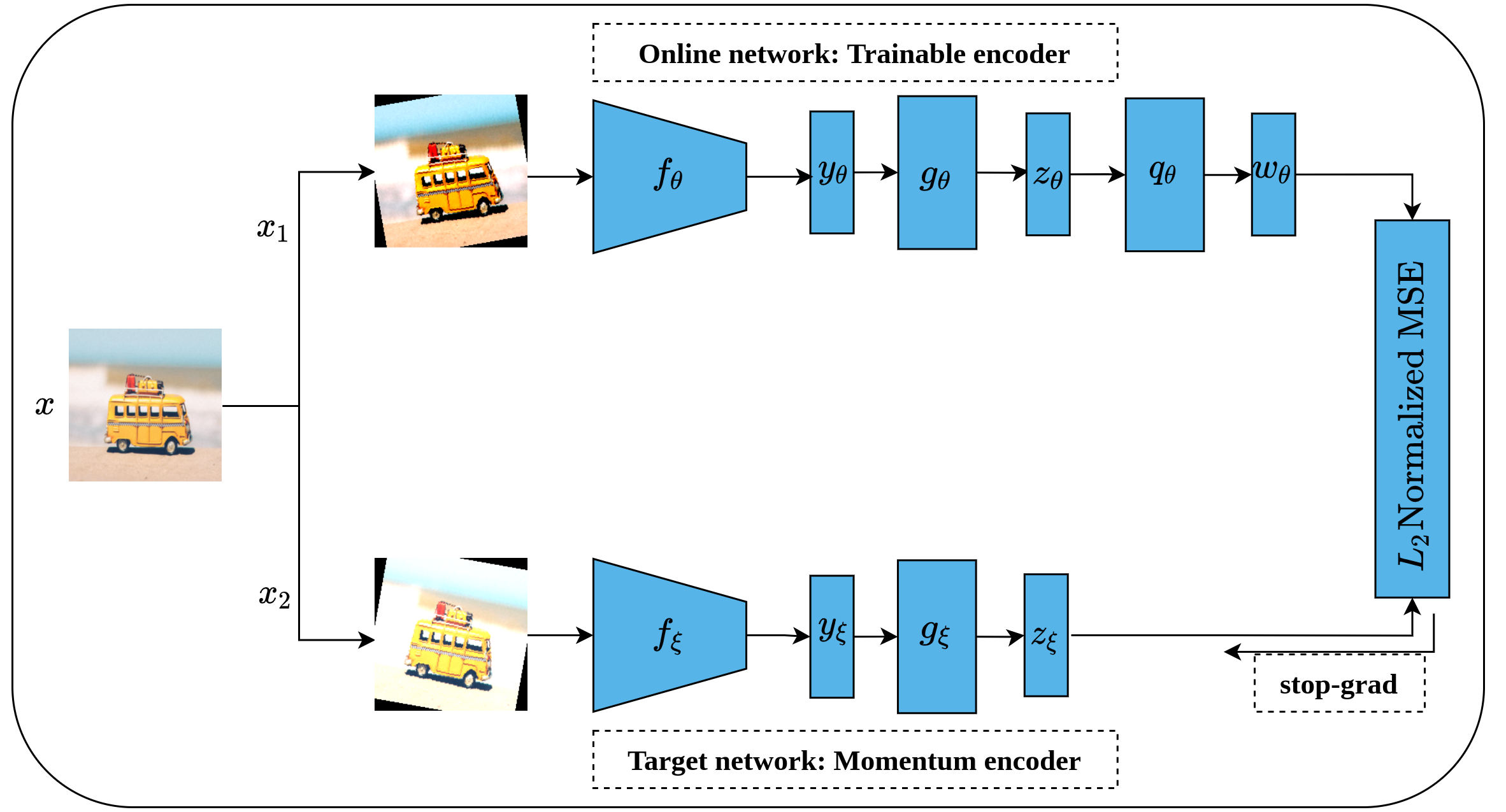}
    \caption{Illustration of BYOL architecture. Image credit: \href{https://www.pexels.com/photo/yellow-van-die-cast-386025/}{Nubia Navarro}.}
    \label{fig16}
\end{figure}

Bootstrap Your Own Latent (BYOL) is an implicit contrastive learning approach proposed by \cite{NEURIPS2020_f3ada80d} that omits the need for negative samples during the training. More clearly, BYOL architecture consists of two networks as shown in Figure \ref{fig16}. The first network is a trainable network, called Online Network, denoted by $\theta$ which consists of a  representation head $f_{\theta}$, a projection head $g_{\theta}$ and a prediction head $q_{\theta}$. The second network is a non-trainable and randomly initialized network, called Target Network, denoted by $\xi$ and has the same architecture as the Online Network except for the prediction head. Target Network acts as a slow-moving average of the Online Network and is updated based on the gradients update in the Online Network via the moving average. To train BYOL architecture, two augmented views $(x_1, x_2)$ are generated from the input image $x$ by applying two different augmentation operations $(t,\acute{t})$. Consequently, both augmented views pass to the two networks for encoding $(y_{\theta}, y_{\xi})$ and representations generation $(z_{\theta}, z_{\xi})$,  while $z_{\theta}$ passes through the prediction head to produce the prediction $w_\theta$ for the subsequent computation. Following that, both  $w_\theta$ and $z_{\xi}$ are normalized via L2 norm and accordingly fed into mean squared error (MSE) loss function for optimization rather than contrastive loss. It is worth noting that the gradients flow back only over the Online Network and stopped for Target Network as indicated in figure \ref{fig16} by the term stop-grad which is updated with the momentum equation as a function of the Online Networks parameters $\theta$. Since the target network acts as the moving average of the online network, the online representations should be predicted of the target representations and vice versa. BYOL can learn semantic features by minimizing similarity metrics between the output of each network. Hence, both networks learn interactively from each other from the same image while omitting the need for negative samples.

\subsubsection*{Swapping assignments between multiple views}

While the previous contrastive methods are instance-discrimination-based methods, Swapping Assignments between multiple views (SwAV) is a cluster-discrimination-based method proposed by \cite{NEURIPS2020_70feb62b}. Two major elements form the core of the SwAV method including a multi-crop augmentation strategy and the online clustering assignment. The multi-crop strategy aims at generating multiple views of the same image without increasing the memory and computational requirements during the training. This is achieved through generating two global views with standard resolution crops, e.g., $224\times224$, and $V$ local views with smaller resolution crops, e.g., $96\times96$. This way, a multi-crop strategy enables producing multiple views rather than just pairs without affecting the computational and memory cost. Besides, each generated view undergoes additional random transformations such as those implemented in SimCLR \citep{chen2020simple}. On the other side, unlike offline clustering assignment methods which require a complete pass over the dataset to compute the clusters' assignment which becomes computationally intensive in the case of large datasets; Online clustering allows computing clusters' assignment by mapping the encoded views to a prototype vector $C$ on the current batch by treating it as an optimal transportation problem (Sinkhorn-Knopp \citep{cuturi2013sinkhorn}).

\begin{figure}[!ht]
    \centering
    \includegraphics[width=11cm, height=5.5cm]{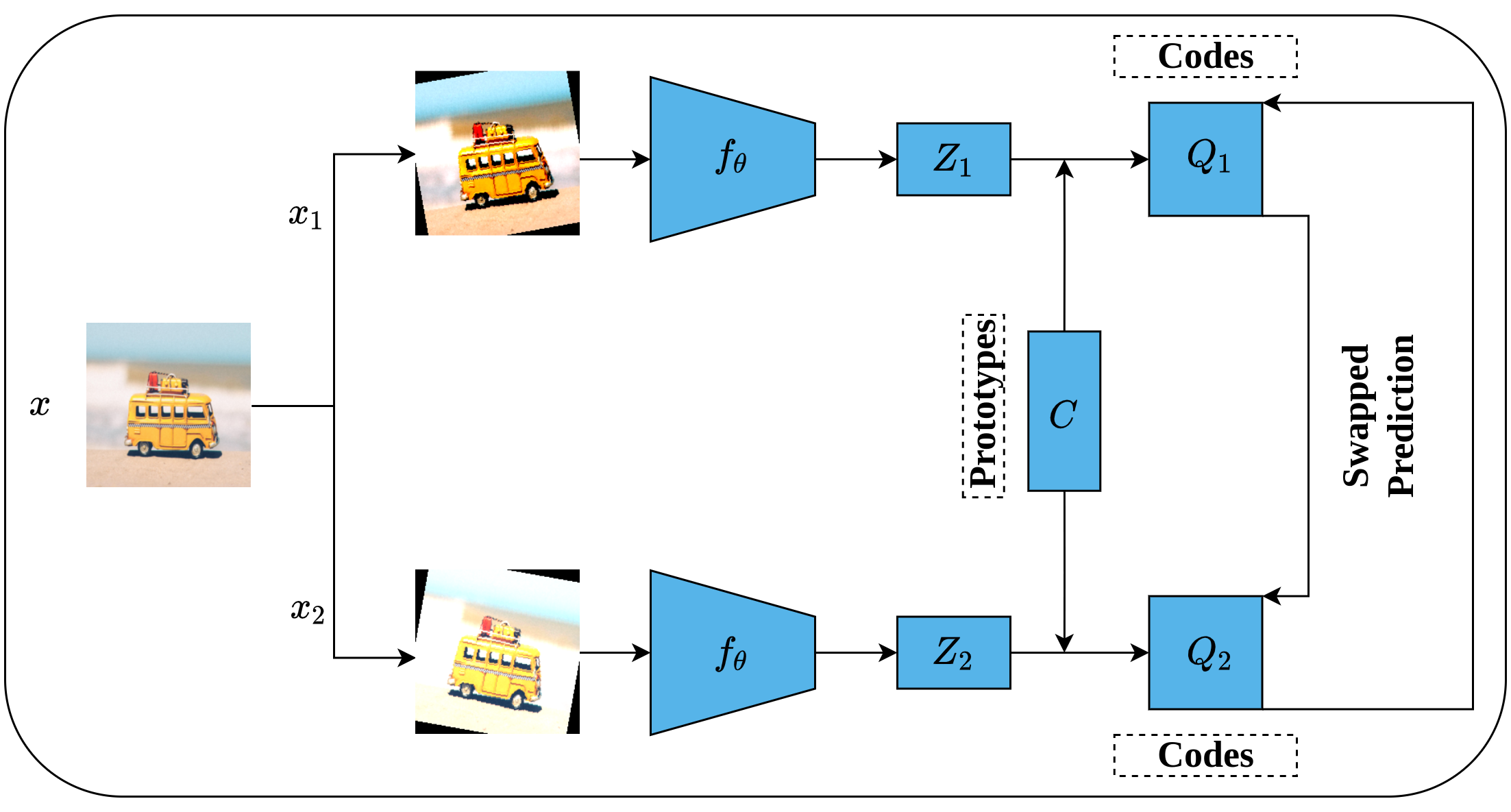}
    \caption{Illustration of SwAV framework. Image credit: \href{https://www.pexels.com/photo/yellow-van-die-cast-386025/}{Nubia Navarro}.}
    \label{fig17}
\end{figure}

Figure \ref{fig17} depicts the complete SwAV architecture. Given an input image $x$, multiple views of the same image are generated by applying a set of random transformation $T$ according to a multi-crop augmentation strategy resulting in $x_{nt}$ augmented views. For simplicity, we will consider one global view $x_1$ and one local view $x_2$. Consequently, the generated views are passed into convolutional encoders $f_{\theta}$, ResNet50 \citep{he2016deep} in SwAV case, followed by two dense layers with ReLU activation \citep{nair2010rectified} to generate feature vectors $(Z_1,Z_2)$. In fact, the initial steps in SwAV do not differ significantly from those of SimCLR \citep{chen2020simple} except in the augmentation strategy. Following that, the feature vectors $(Z_1,Z_2)$ are passed through a dense layer with linear activation called prototype layer $C$ which is responsible for mapping the feature vectors into $K$ learnable prototypes (clusters) grouped in a matrix such that $C = [c_1 , c_2,...., c_K]$. It is worth noting that $K$ value is not inferred but user-defined, while the $C$ values represent the weights matrix of the prototype layer. To compute clusters assignments online, only the features of the current batch are used where Sinkhorn-Knopp algorithm is employed to generate the cluster assignments (codes) $(Q_1,Q_2)$ that represent the mapping of feature vectors into clusters in a way that maximizes the similarity between them. Further, Sinkhorn-Knopp enforces the equipartition constraint which prevents assigning all features into a single cluster. Eventually, a swapped prediction problem is performed upon codes generation. Intuitively, given two different views of the same image, they should maintain similar information. Therefore, it is possible to predict the codes of one view from the features vector of the other. This is achieved by minimizing the cross-entropy loss between the code of one view and the softmax of the similarity of the features vector to all clusters. This way, SwAV takes the advantage of contrasting clusters of data with similar features rather than performing pair-wise comparison over the whole training sets as seen in the previous methods.

To sum up, we opted to provide a high-level overview of each of the previously discussed methods as this article is intended for self-supervised applications in medical imaging which renders it prone to nonspecialist readers from the medical field. One more point to mention is that despite the fact that these methods are developed on natural images, they can be transferred to the medical imaging field as we will see in the next section. Such property encouraged us to briefly discuss them before proceeding with the application of self-supervised learning in medical imaging. Table \ref{table1} summarizes the discussed pretext tasks according to their categories, while table \ref{table25} provides an access to the code repository of these works which is provided in appendix A. 

\begin{table}[ht]
	\centering
	\begin{tabular}{l|l|l|l}

		No. & Authors     & Category     & Method  \\
		\hline
		   
		   1&\cite{dosovitskiy2015discriminative}  & Predictive  & Exemplar CNN \\
		   
		   2&\cite{doersch2015unsupervised}  & Predictive  & Relative position prediction  \\
		    
		    3&\cite{noroozi2016unsupervised} & Predictive &  Jigsaw puzzle  \\
		    
		    4&\cite{komodakis2018unsupervised} & Predictive & Rotation prediction \\

		    5&\cite{vincent2008extracting} & Generative & Denoising auto-encoder \\
		    
		    6&\cite{pathak2016context} & Generative &  Image inpainting \\
		    
		    7&\cite{zhang2016colorful} & Generative & Image colorization  \\
		    
		    8&\cite{zhang2017split} & Generative & Split-brain auto-encoder \\
		    
		    9&\cite{DBLP:unsupjournals/corr/RadfordMC15} & Generative & Deep Convolutional GAN \\
		    
		    10&\cite{donahue2016adversarial} & Generative & Bi-directional GAN \\

		    11&\cite{DBLP:cpcjournals/corr/abs-1807-03748} & Contrastive & CPC \\

		    12&\cite{he2020momentum} & Contrastive & MoCo  \\
            
		    13&\cite{chen2020simple} & Contrastive &  SimCLR \\
		    
		    14&\cite{NEURIPS2020_f3ada80d} & Contrastive &  BYOL \\
		    
		    15&\cite{NEURIPS2020_70feb62b} & Contrastive & SwAV \\

	\end{tabular}
	\caption{Summary of self-supervised learning pretext tasks.}
	\label{table1}
\end{table}

\subsection*{Resources in self-supervised learning}

We provided a curated list of pretext tasks that acted as milestones in the history of self-supervised learning in the computer vision field, however, the efforts in this research area are not limited to those methods. As a result, we developed a list of self-supervised learning resources that includes review articles, surveys, and papers as shown in Table \ref{table2} for those readers who need to enhance their understanding of the field. For in-depth reviews about self-supervised learning, we highly recommend the readers to refer to one of the following articles: \citet{jing2020self} provided an extensive review of self-supervised learning methods for visual features learning from image and video data, and \cite{ohri2021review} provided a comprehensive review and performance comparison for a large list of the most recent self-supervised learning approaches developed for image data. Further, \citet{schmarje2021survey}  reviewed various deep learning methods for image classification with fewer labels where self-supervised learning is one of their work dimensions. For Contrastive learning, both \citet{le2020contrastive} and \citet{jaiswal2021survey} provided a comprehensive survey on contrastive self-supervised methods for different research areas such as computer vision and natural language processing. \citet{liu2021self} summarized a set of generative and contrastive self-supervised learning approaches from computer vision, natural language processing, and graph learning. To access these lists of papers, readers may visit the following two repositories: Awesome-self-supervised-learning\footnote{https://github.com/jason718/awesome-self-supervised-learning} which covers a curated list of research articles for self-supervised learning from different research areas. In addition, Awesome-contrastive-learning\footnote{https://github.com/asheeshcric/awesome-contrastive-self-supervised-learning} is a curated list of papers that is mainly dedicated to contrastive learning methods.

\begin{table}[ht]
	\centering
	\begin{tabular}{l|l|l|l|l}

		No.&Authors  & Type & Title  & Venue  \\
		\hline
		   
		  1&\citet{jing2020self} & Survey & \vtop{\hbox{\strut Self-supervised visual feature learning}\hbox{\strut with deep neural networks: A survey}} & \vtop{\hbox{\strut IEEE Transactions on}\hbox{\strut Pattern Analysis and}\hbox{\strut Machine Intelligence }}  \\

		  2&\citet{ohri2021review}& Review & \vtop{\hbox{\strut Review on self-supervised image }\hbox{\strut  recognition using deep neural}\hbox{\strut networks }} & \vtop{\hbox{\strut Knowledge-Based}\hbox{\strut Systems }}  \\

		  3&\citet{schmarje2021survey}& Survey & \vtop{\hbox{\strut A survey on semi-, self- and }\hbox{\strut unsupervised learning in image} \hbox{\strut classification }} & \vtop{\hbox{\strut IEEE Acsess }\hbox{\strut  }}  \\

		  4&\citet{le2020contrastive}& Review & \vtop{\hbox{\strut Contrastive representation learning: }\hbox{\strut A framework and review}} & \vtop{\hbox{\strut IEEE Access }\hbox{\strut  }}  \\

		  5&\citet{liu2021self}& Review & \vtop{\hbox{\strut Self-supervised learning: }\hbox{\strut Generative or contrastive }} & \vtop{\hbox{\strut IEEE Transactions on  }\hbox{\strut Knowledge and Data}\hbox{\strut Engineering }}  \\

		  6&\citet{jaiswal2021survey}& Survey & \vtop{\hbox{\strut  Survey on contrastive self-supervised  }\hbox{\strut learning}} & \vtop{\hbox{\strut Technologies }\hbox{\strut  }}  \\

		  7&Jason Ren & Papers list & \vtop{\hbox{\strut Awesome self-supervised }\hbox{\strut learning }} & \href{https://github.com/jason718/awesome-self-supervised-learning}{Github}  \\

		  8&Ashish Jaiswal & Papers list &   \vtop{\hbox{\strut Awesome contrastive learning }\hbox{\strut }} & \href{https://github.com/asheeshcric/awesome-contrastive-self-supervised-learning}{Github}  \\

	\end{tabular}
	\caption{A summary of Self-supervised learning resources}
	\label{table2}
\end{table}

\section*{Self-supervised methods in medical imaging}

Mainly, there are two paths to follow when employing self-supervised learning in medical images analysis \citep{chen2021recent}. The first path is to directly adopt one of the pre-designed pretext tasks from the computer vision field as given in Section 3 or alternatively develop modified versions of these tasks and employ them in the medical context. On the other side, the second path exploits knowledge from the medical domain and computer vision to design novel pretext tasks that target medical images. We prefer to categorize self-supervised learning methods, which are used in the medical field, in the same way we categorized self-supervised learning methods in the computer vision field. Further,  after exploring self-supervised learning literature in medical imaging, we discovered that some researchers tend to utilize multiple methods separately or collectively in a multi-tasking fashion. So, we added an additional category called multiple-tasks/multi-tasking to fit such works.

\subsection*{Predictive methods in medical imaging}

Inspired by relative position prediction \citep{doersch2015unsupervised} task, \cite{zhang2017self} introduced slices ordering pretext task. Knowing that 3D medical images such as CT and MRI scans can be represented as a group of successive 2D slices, such property can be used as an auxiliary supervision signal to learn a good representation. As a result, the authors treated the slice ordering task as a binary classification problem by developing a Siamese convolutional architecture called Paired-CNN that receives two successive slices and predicts their spatial order as below or above. The authors tested their proposed task on fine-grained body part recognition (regression) as a downstream task.

\cite{spitzer2018improving} proposed to predict the geodesic distance between two patches located on the brain surface to learn a rich representation of the human brain. They trained a Siamese architecture with two identical branches and weights sharing to accomplish this task. The defined distance between two patches is the Euclidean distance while the ground truth distance is computed manually from the input data. Besides the distance prediction, the authors included the 3D location coordinates prediction of the input patches to the same task which improved the accuracy and convergence of the predicted distances. Lastly, their approach was evaluated on Cytoarchitectonic segmentation as a downstream task.

\cite{bai2019self} proposed anatomical position prediction pretext task from cardiac MRI scans for segmentation purposes. As the cardiac MRI scans provide several cardiac views from different orientations, e.g., short-axis, 2CH long-axis, and 4CH long-axis. Different cardiac anatomical regions, e.g., left and right atrium and ventricle can be expressed using these views. Such properties motivated the authors to define a set of anatomical positions with respect to a certain view as bounding boxes and forced the network to predict these anatomical positions. For the downstream task, a private dataset of 200 annotated cardiac MRI scans was used for evaluation purposes.

\cite{li2020self} employed self-supervised learning to improve the pseudo-labeling uncertainty estimation in semi-supervised medical images' segmentation by proposing a novel methodology called self-loop uncertainty. They adopted the Jigsaw puzzle pretext task \citep{noroozi2016unsupervised} in their approach and introduced random patches' rotation with angles of [$0^{\circ}$, $90^{\circ}$, $180^{\circ}$, $270^{\circ}$] to secure learning translation and rotation invariant features. Further, they omitted the need for Siamese architecture as compared to the original Jigsaw puzzle by combining the input patches into a single image for subsequent permutation classification. Besides the labeled data, they leveraged unlabeled data for uncertainty estimation in semi-supervised settings. Two different segmentation tasks were considered for validating the methodology including nuclei segmentation and skin lesion segmentation as down-stream tasks. 

\cite{taleb2021multimodal} presented another work that is inspired by Jigsaw puzzle-solving \citep{noroozi2016unsupervised} that exploits multiple imaging modalities, e.g.: T1 and T2 scans simultaneously called multi-modal Jigsaw puzzle. A significant improvement has been brought to the original Jigsaw puzzle, besides the multi-modal settings, represented by the employment of Sinkhorn network \citep{mena2018learning} for Jigsaw puzzle solving. Sinkhorn network utilizes the Sinkhorn operator as an alternative to the Softmax function which in turn enables solving the Jigsaw puzzle by learning a permutation task rather than a classification task. They also introduced cross-modal synthesis data generation using CycleGAN architecture \citep{zhu2017unpaired} to increase the amounts of data available for self-supervision. On the downstream side, four tasks were utilized for method validation including brain tumor segmentation, prostate segmentation, liver segmentation, and survival days prediction (regression).

\cite{zhuang2019self} proposed a novel pretext task that is inspired by the early work of \cite{noroozi2016unsupervised} on Jigsaw puzzle solving for 3D medical data called Rubik cube recovery. Two operations constitute the Rubik cube recovery pretext task including cube rearrangement and cube rotation. The same logic of the original Jigsaw puzzle task is adopted in the Rubik cube recovery task with 3D input as a Rubik cube partitioned into a 3D grid of $2\times2\times2$ sub-cubes rather than 2D input with respect to the rearrangement task. To introduce additional complexity, the authors introduced the cube rotation process and limited it to only $180^{\circ}$ vertically and horizontally. This way, the authors secured learning translation and rotation invariant features as opposed to the original Jigsaw puzzle task which secures learning only translation-invariant features. Two downstream tasks were used for evaluation purposes including brain hemorrhage classification and brain tumor segmentation which showed competitive performance.

As an extension of the previous work, \cite{zhu2020rubik} introduced the Rubik cube+ pretext task which adds an additional level of complexity to the Rubik cube recovery problem represented as cube masking identification on the top of both cube rearrangement and cube rotation. Masking identification operation can be viewed as randomly blocking part of the information in a certain cube by masking. The intuition behind masking identification is that robust features learning can be achieved by solving harder tasks. Rubik cube+ was evaluated on the same downstream tasks from the previous work which showed slight improvement.

\cite{nguyen2020self} proposed spatial awareness pretext task that is able to learn semantic and spatial representations from volumetric medical images. Spatial awareness is inspired by the context restoration framework \citep{chen2019self} but was treated as a classification problem. For a certain 3D image, a single slice is selected in addition to a neighboring slice in the range $[-2,2]$ where this range represents the spatial index. Following that, two patches of predefined dimensions are selected randomly and swapped between the two slices T times. Lastly, a classification network is trained to predict if the input slice is corrupted or not to learn semantic representations. Further, the network is trained to learn the spatial index which enables learning spatial features.

Table \ref{table3} summarizes the predictive self-supervised learning methods in medical imaging.

\begin{table}[!ht]
	\centering
	\begin{tabular}{l|l|l|l}

		No. &Authors     & Pretext task & Down-stream task  \\
		\hline
		   
		1&\cite{zhang2017self} &Slices ordering &  Body parts recognition \\

		2&\cite{spitzer2018improving} & Geodesic  distance prediction & Brain area segmentation  \\

		3&\cite{bai2019self}    & Anatomical position prediction &  \vtop{\hbox{\strut Short-axis cardiac MRI segmentation}\hbox{\strut  long-axis cardiac MRI segmentation}}  \\

		 4&\cite{li2020self} & Jigsaw puzzle  &\vtop{\hbox{\strut Nuclei Segmentation	    }\hbox{\strut Skin lesions segmentation }} \\

		 5&\cite{taleb2021multimodal} & Jigsaw puzzle  & \vtop{\hbox{\strut  Brain tumor segmentation  }\hbox{\strut Liver segmentation }\hbox{\strut Prostate segmentation }} \\

		6&\cite{zhuang2019self} & Rubik cube &  \vtop{\hbox{\strut Brain tumor segmentation}\hbox{\strut  Brain hemorrhage classification}} \\

		7&\cite{zhu2020rubik}    & Rubik cube+ &  \vtop{\hbox{\strut Brain tumor segmentation}\hbox{\strut  Brain hemorrhage classification}}  \\

		8&\cite{nguyen2020self}    & Spatial awareness &  \vtop{\hbox{\strut Organ at risk segmentation}\hbox{\strut  Intracranial Hemorrhage detection}} \\

	\end{tabular}
	\caption{Summary of predictive self-supervised learning methods in medical imaging}
	\label{table3}
\end{table}

\subsection*{Generative methods in medical imaging}

\cite{ross2018exploiting} adopted image colorization pretext task \citep{zhang2016colorful} for solving endoscopic medical instruments segmentation task from endoscopic video data. The authors did not utilize the original architecture as in the colorization task, but instead, a conditional GAN architecture was employed to encourage generating more realistic colored images, while six datasets from medical and natural domains were used in the evaluation of downstream tasks.

\cite{chen2019self} proposed a novel generative pretext task, called context restoration, that is inspired by the early works of relative position prediction \citep{doersch2015unsupervised} and context encoder \citep{pathak2016context}. The authors described the context restoration task as a simple and straightforward method in which two isolated patches are selected randomly and their positions are swapped. The swapping process repeats itself iteratively to produce a corrupted version of the input image but preserves the input image's overall distribution. Following that, a generative model is employed to restore the corrupted image to its original version. Three downstream tasks were used to test the context restoration feasibility including fetal standard scan plane classification, abdominal multi-organ localization, and brain tumor segmentation. 

Another work that is built on the same idea of context restoration is called Models Genesis and is performed by \cite{zhou2019models} for 3D medical images. As opposed to the context restoration pretext task, models genesis introduced four distortion operations, namely, non-linear transformation using the Bézier transformation function, local pixel shuffling which is similar to the swapping operation in context restoration but in 3D settings, in-painting which is similar to context encoder method and out-painting which is the inverse operation of in-painting. It is worth noting that each input volume undergoes the first two operations and only one of the remaining operations. Consequently, a generative model is built to restore the distorted image to its original context. Six downstream tasks were used to evaluate their method in terms of segmentation and classification tasks.

\cite{matzkin2020self} designed a self-supervised approach for bone flab reconstruction that results from decompressing craniectomy (DC) operations using normal CT scans rather than DC post-operative annotated CT scans. DC is the surgical procedure of removing part of the skull due to different causes such as stroke and traumatic brain injury. The authors designed a virtual craniectomy approach to simulate the DC from normal CT scans that generate DC post-operative CT scans with bone flaps removed from different parts of the upper head which in turn serve as input for the reconstruction model. Consequently, two strategies were proposed to reconstruct the bone flab including direct estimation as well as reconstruction and subtraction. Further, two architectures were employed including U-Net \citep{ronneberger2015u} and denoising auto-encoder \citep{vincent2008extracting}.

\cite{hervella2020learning} proposed a multi-modal reconstruction task as a self-supervised approach for retinal anatomy learning. The main assumption is that different modalities for the same organ can provide complementary information which enables learning useful representations for the subsequent tasks. The authors proposed to reconstruct fundus fluorescein angiography photos from color fundus photos using aligned pairs from both modalities for the same patient's eye. Further, U-net architecture \citep{ronneberger2015u} is employed for the sake of completing the reconstruction task along with structural similarity index map (SSIM) \citep{wang2004image} as a loss function. Subsequent research by the same authors experimented with their approach on different ophthalmic oriented down-stream tasks such as retinal vascular segmentation \citep{MoranoHBNR20}, joint optic disc and cup segmentation \citep{hervella2020multi} and diagnosis of retinal diseases \citep{hervella2021self}.

\cite{holmberg2020self} suggested that designing an effective pretext task for medical domains must accurately extract disease-related features which are typically present in a small part of the medical image. Hence, such condition makes traditional pretext tasks that are dominated by the presence of larger objects in natural images ineffective for the medical context. As a result, they developed a novel pretext task for ophthalmic diseases diagnosis called cross-modal self-supervised retinal thickness prediction that employs two different modalities including optical coherence tomography scans (OCT) and infrared fundus images. Initially, retinal thickness maps are extracted from OCT scans by developing a segmentation model using a small annotated dataset which then serves as ground-truth labels for the actual pretext task. Following that, a model is trained to predict the thickness maps using unlabeled infrared fundus images and the predicted thickness map from the previous step as labels. Learning disease-related features using the proposed approach has been validated by three experienced ophthalmologists. Further, the quality of their task was assessed on diabetic retinopathy grading using color fundus as a downstream task.

\cite{prakash2020leveraging} adopted image denoising approach as a pretext task for nuclei images' segmentation. A special denoising architecture called Noise2Void \citep{krull2019noise2void} was employed as a self-supervised pretraining method. Further, four scenarios are evaluated for segmenting nuclei images including random initialization with noisy images,  random initialization with denoised images, fine-tuning with noisy images, and fine-tuning with denoised images.  The results showed the superiority of self-supervised denoising as opposed to the random initialization.

\cite{hu2020self} adopted context encoder framework \citep{pathak2016context} as a pretext task along with DICOM meta-data as a weak supervision method to learn robust representations from ultrasound imaging. On the top of the context encoder, the authors introduced additional projection discriminator \citep{miyato2018cgans,luvcic2019high} network that produces a feature vector of the inpainted image which to be fed into the classification head and projection head. The classification head classifies the context encoder output as real or fake; while the projection head acts as a conditional classifier that incorporates the DICOM meta-data as weak labels. For DICOM meta-data, two tags were employed including the prop type and the study description as they directly relate to the ultrasound semantic context.

Another extension to Rubik cube pretext tasks is performed by \cite{tao2020revisiting} as Rubik cube++ which introduced two substantial changes to the original Rubik cube problem. On the first hand, they introduced the concept of volume-wise transformation which bounds the sub-cubes rotation operation to the neighboring sub-cubes as in playing a real Rubik cube game and as opposed to \cite{zhuang2019self} where the sub-cubes are rotated individually. On the second hand, rather than treating the Rubik cube as a classification problem, it has been treated as a generative problem using GAN-based architecture where the generator's role is to restore the original order of the Rubik cube before applying the transformation, while the discriminator role is to discriminate between the correct and wrong arrangement of the generated cubes. As a downstream task, Rubik cube++ has been tested on two segmentation tasks including pancreas segmentation and brain tissues segmentation.

Table \ref{table4} summarizes the generative self-supervised learning methods in medical imaging.

\begin{table}[!ht]

	\centering
	\begin{tabular}{l|l|l|l}

		No.&Authors     & Pretext task & Down-stream task \\
		\hline

		1&\cite{ross2018exploiting} & Image Colorization &  Surgical instruments segmentation \\

	    2&\cite{chen2019self} & Context restoration  &
		   \vtop{\hbox{\strut Fetal image classification}\hbox{\strut Abdominal multi-organ localization}\hbox{\strut Brain tumour segmentation}} \\

		     3&\cite{zhou2019models} &  Models Genesis &
		   \vtop{\hbox{\strut Lung nodule segmentation}\hbox{\strut FPR for nodule detection}\hbox{\strut FPR for pulmonary embolism }\hbox{\strut Liver segmentation}\hbox{\strut pulmonary diseases classification}\hbox{\strut RoI, bulb, and background classification}\hbox{\strut Brain tumor segmentation}} \\

		  4&\cite{matzkin2020self}    & Skull reconstruction&  Bone flap volume estimation\\

		  5&\cite{hervella2020learning} & Multi-modal reconstruction &  \vtop{\hbox{\strut Fovea localization}\hbox{\strut  Optic disc localization}\hbox{\strut  Vasculature segmentation}\hbox{\strut  Optic disc segmentation}} \\

        6&\cite{holmberg2020self} &  Cross modal retinal thickness prediction & Diabetic retinopathy grading \\

		 7&\cite{prakash2020leveraging} & Image denoising & Nuclei images segmentation \\

		 8&\cite{hu2020self} & Context encoder & \vtop{\hbox{\strut Quality score classification }\hbox{\strut Thyroid nodule segmentation}\hbox{\strut  Liver and kidney segmentation}} \\

		  9&\cite{tao2020revisiting} & Rubik cube++ &  \vtop{\hbox{\strut Pancreas segmentation}\hbox{\strut  Brain tissue segmentation}}  \\

	\end{tabular}
		\caption{Summary of generative self-supervised learning methods in medical imaging.}
	\label{table4}
\end{table}

\subsection*{Contrastive learning in medical imaging}

\cite{jamaludin2017self} exploited longitudinal spinal MRI scans as a self-supervised contrastive learning task. This is supported by the fact that time-separated scans from the same patient will share similar representations. As a result, they trained a Siamese network that contrasts two vertebral bodies (VB) MRI scans separated by a period of time. regardless of whether the two images belong to the same patient or not by employing a contrastive loss function \cite{chopra2005learning}. Along with the contrastive loss, they employed a categorical cross-entropy loss to classify the VB scans into seven classes T1-S1. For the downstream task, they tested the pre-trained model on the disc degeneration grading task which showed superior performance in comparison to the random initialization. 

\cite{lu2020@semi} adopted contrastive predictive coding \citep{DBLP:cpcjournals/corr/abs-1807-03748} along with multiple instance learning (MIL) \citep{ilse2018attention} for the classification of breast cancer histology images. As a first stage, CPC is employed to learn rich representations from breast cancer histopathological images rather than learning features from scratch using the MIL network. The results showed superior performance as compared to training from scratch and the ImageNet pre-trained model.

Contrastive predictive coding \citep{DBLP:cpcjournals/corr/abs-1807-03748} is originally designed for 2D data, \cite{zhu2020embedding} extended the early work on contrastive predictive coding in a way that enables handling 3D data by developing a new method called Task-related CPC. Initially, supervoxels are generated using a simple linear iterative clustering method (SLIC) \citep{achanta2012slic} from the input volume to detect the potential lesion areas. Consequently, the sub-volumes that surround the generated supervoxels are cropped to act as the input of the TCPC encoder network. A U-shape path around the center of the generated supervoxel is employed as compared to the original CPC which employs a straight path to achieve better characterization of the lesion. Further, the recurrent neural network acts as the auto-regressor which generates the future predictions, while the whole architecture is optimized using the InfoNCE loss \citep{DBLP:cpcjournals/corr/abs-1807-03748}. Brain hemorrhage classification and lung nodule classification tasks were utilized as downstream tasks.

\cite{xie2020pgl} stated that self-supervised approaches, in general, and contrastive approaches, in specific, are known to consider the global consistency of the input data while ignoring the local consistency. The authors introduced the Prior-Guided Local (PGL) algorithm for 3D medical images' segmentation which extended the early work on the BYOL method \citep{NEURIPS2020_f3ada80d} to consider the local consistency between the different views of the same region. To achieve this, an additional block, called a prior-guided aligner, is added on top of the projection head for both online and target networks used in the original BYOL architecture. The role of the prior-guided aligner is to exploit the augmentation information applied to the input image prior to guide-aligning the features extracted from different views of the same region. Lastly, a local consistency loss function is employed to minimize the difference between the aligned local features. Four downstream segmentation tasks were employed for evaluation purposes, including liver tumors, kidney tumors, spleen, and abdominal organs.

\cite{li2020self-sup} proposed patient's feature-based Softmax embedding loss function to learn modality and transformation invariant features as well as patients' similarity features using ophthalmic data in contrastive settings. Modality invariant features are learned by combining color fundus photos with a synthesized fundus fluorescein angiography photo of the former photo, while transformations invariant is represented by the ordinary augmentation techniques of the color fundus photo. Such triplet of photos is assumed to share similar features for the same patient. Consequently, to learn patients' similarity features, the triplet of each patient image is considered as a contrasting basis where the features of the same patients are pulled together, while features from other patients are pushed apart using the proposed loss function.

\cite{pmlr-v143-sowrirajan21a} adopted MoCo  \citep{he2020momentum} approach to build self-supervised pre-trained models for chest X-Ray classification problem. They used pre-trained models on ImageNet \citep{deng2009imagenet} in a supervised fashion as initialization weights for the self-supervised training to speed up the convergence. Further, they suggested that not all augmentation strategies implemented in the original MoCo paper can fit into gray-scale images. Instead, they settle only to use  random partial rotation and horizontal flipping. In addition, they tested their work on an external chest X-Ray dataset to examine the generalizability of their work on tasks from the same domain, which showed the possibility of transferring the self-supervised learned knowledge to other related tasks.

\cite{vu2021medaug} proposed the MedAug approach as an augmentation strategy that benefits from the patient meta-data when training MoCo framework \citep{he2020momentum} as an extension of the early work performed by \cite{pmlr-v143-sowrirajan21a}. More clearly, MedAug requires that the different views must come from the same patient, as such images are expected to be rich in pathological features. In addition, MedAug considers studying number and laterality as two additional conditions derived from the patient meta-data. For the same patient, the study number represents images taken in different sessions, while laterality represents the orientation as frontal or lateral. This way, MedAug leveraged medical knowledge to the learning algorithm rather than depending merely on the transformations obtained by ordinary augmentation techniques to generate positive views. MedAug was tested on pleural effusion classification from chest X-ray as a downstream task.

\cite{sriram2021covid} purely adopted MoCo \citep{he2020momentum} as an approach for COVID patients deterioration prediction tasks. They used non-COVID chest X-ray images from different public datasets to train MoCo for the subsequent tasks. On the other side, the authors defined three prediction tasks that indicate COVID patient deterioration including single image prediction, oxygen requirements prediction, and multiple image prediction as downstream tasks. The first two tasks are ordinary classification problems from a single image;  while the third one requires multiple time-indexed radiographs. A continuous positional embedding module was employed to obtain representations from a set of time-indexed radiographs. 

Another similar work performed by \cite{chen2021momentum}, which adopted MoCo as a pretraining method, uses chest CT scans for COVID diagnosis via a few-shot learning prototypical network \citep{NIPS2017_cb8da676} as a down-stream task. Similarly, public non-COVID chest CT was utilized for MoCo training; and two public COVID datasets were utilized for evaluation.

\cite{NEURIPS2020_949686ec} provided two significant improvements to the SimCLR \citep{chen2020simple} contrastive learning approach for 3D images segmentation by developing domain-specific and problem-specific knowledge simultaneously. To elaborate more on the domain-specific knowledge, the original contrastive loss (NT-Xent) maximizes the similarity between a pair of transformed versions of the input image by augmentation alone to obtain a global representation. Also, 3D medical images consist of a set of sequential images that depict similar anatomical regions. Hence, such sequences can be exploited as a positive pair to learn a global representation. On the other side for problem-specific knowledge, a segmentation task that is considered a pixel-wise prediction problem requires local representation. As a result, the authors introduced a local contrastive loss function that helps learn a local representation based on the similarity between the local regions within the input volume. It is worth noting that the proposed approach employs encoder-decoder architecture, where the encoder is optimized with global loss while the decoder is optimized with the local loss. Further, cardiac segmentation and prostate segmentation were employed as downstream tasks.

\cite{azizi2021big} adopted a self-supervised contrastive learning approach in a medical context in a way that combines learning features from both unlabelled natural images and unlabelled medical images in a sequential fashion. To elaborate more, they adopted SimCLR \citep{chen2020simple} and introduced novel contrastive learning called Multi-Instance Contrastive Learning (MICLe) which is built on the same logic of SimCLR with minor modifications. The main idea behind MICLe is to leverage the availability of multiple views of a certain pathology from the same patient as the foundation for contrastive learning. Such correlated views of the same patient are considered as positive pairs rather than generating multiple views from the same image as in SimCLR. In their experiments, the authors tested SimCLR on chest X-Ray images dataset with fourteen classes, while MICLe was tested on a Dermatology dataset with twenty-seven classes as a downstream task. 

Table \ref{table5} summarizes the contrastive self-supervised learning methods in medical imaging.

\begin{table}[!ht]
	\centering
	\begin{tabular}{l|l|l|l}

		No. & Authors     & Pretext task & Down-stream task \\
		\hline
		1&\cite{jamaludin2017self} &Longitudinal spinal MRI &   Disc degeneration grading   \\

		 2&\cite{lu2020@semi} & CPC &  Breast cancer classification  \\

		3&\cite{zhu2020embedding} & TCPC & \vtop{\hbox{\strut  Brain hemorrhage classification }\hbox{\strut Lung Nodule classification }} \\

		4&\cite{xie2020pgl} & BYOL &  \vtop{\hbox{\strut Liver segmentation }\hbox{\strut Spleen segmentation }\hbox{\strut Kidney tumour seg. }\hbox{\strut Abdominal organs seg. }} \\

		5&\cite{li2020self-sup} &  Feature-based softmax embedding &
		   \vtop{\hbox{\strut PM classification}\hbox{\strut  AMD classification}\hbox{\strut  Diabetic retinopathy detection}}  \\

		6&\cite{pmlr-v143-sowrirajan21a} & MoCo & \vtop{\hbox{\strut  Tuberculosis detection }\hbox{\strut Pleural effusion classification}} \\

		7&\cite{vu2021medaug} & MoCo & Pleural effusion classification \\

		 8&\cite{sriram2021covid} & MoCo & COVID patient prognosis \\

		 9&\cite{chen2021momentum} & MoCo &  COVID few-shot classification\\

		10&\cite{NEURIPS2020_949686ec} & SimCLR  & \vtop{\hbox{\strut  Cardiac segmentation }\hbox{\strut Prostate segmentation }} \\

		  11&\cite{azizi2021big} & SimCLR  & \vtop{\hbox{\strut  Chest X-ray classification }\hbox{\strut Skin lesions classification }} \\

	\end{tabular}
	\caption{Summary of contrastive self-supervised learning methods in medical imaging.}
	\label{table5}
\end{table}

\subsection*{Multiple-tasks/Multi-tasking in medical imaging}

\cite{tajbakhsh2019surrogate} experimented with three pretext tasks, namely, rotation prediction \citep{komodakis2018unsupervised}, image colorization \citep{larsson2017colorization} and 3D Patch reconstruction \citep{arjovsky2017wasserstein} on three different medical image analysis tasks including False Positive Reduction for nodule detection in chest CT scans, diabetic retinopathy severity classification, lung lobe segmentation and skin segmentation. Due to the substantial differences among the utilized imaging modalities, each of them was assigned a specific pretext task. More clearly, image rotation was employed for both lung lobe segmentation and diabetic retinopathy classification, while colorization was employed for the skin segmentation task and finally 3D patch reconstruction was employed for nodule detection. 

\cite{jiao2020self} proposed temporal order correction and spatio-temporal transformation prediction pretext tasks to learn good representations from fetal ultrasound videos. For the first task, the order of the ultrasound video frames is shuffled and the role of the task is to predict the correct order of the shuffled frames. For the second task, certain affine transformations are applied to the input video and the role of the task is to predict the applied transformations. To train both tasks jointly, the authors proposed two strategies including a Siamese network with partial weights sharing that learns two tasks simultaneously with one branch for each task. The second strategy is called objective disentanglement which enables incorporating the proposed task into the same input video and training the network to recognize both of them.

\cite{li2020multi} combined two colorization-based pretext tasks into a single multi-tasking framework called ColorMe to learn useful representations from scopy images. In a similar way to the original colorization task \citep{zhang2016colorful}, the authors proposed to predict red and blue channels from the green channel in an RGB scopy images to obtain local features. On the other side, the authors proposed color distribution estimation of the red and blue channels to force learning of global features. Then, both tasks are trained jointly and evaluated on two downstream tasks, namely, cervix type classification and skin lesion segmentation.

\cite{NEURIPS2020_d2dc6368} suggested that rich representations can be learned from medical images with 3D nature instead of 2D images. For this reason, they applied five pre-designed pretext tasks, namely, CPC \citep{DBLP:cpcjournals/corr/abs-1807-03748}, exemplar CNN \citep{dosovitskiy2015discriminative}, rotation prediction \citep{komodakis2018unsupervised}, relative position prediction \citep{doersch2015unsupervised} and Jigsaw puzzle \citep{noroozi2016unsupervised} to be adaptive with medical images of 3D nature. Their methods were tested on two 3D downstream tasks which are brain tumor segmentation and pancreas tumor segmentation. 

\cite{luo2020retinal} proposed a self-supervised fuzzy clustering network as a pretext task for color fundus photo classification. The proposed approach consists of auto-encoder architecture which is responsible for initial features learning from the input data as a first stage. In addition, a clustering module that guides the self-supervision process is employed as a second stage. After gaining the initial representations, the Fuzzy C-means algorithm is utilized  \citep{bezdek1984fcm} on top of the encoder network to cluster similar inputs into predefined clusters and update the encoder weights accordingly. The learned weights, after the clustering phase is complete, are transferred to the downstream task.

\cite{haghighi2020learning} introduced Semantic Genesis as an extension to the previous work on Models Genesis framework \citep{zhou2019models}. Besides features learning by restoration, the authors introduced two additional functionalities called self-discovery and self-classification. Self-discovery is the first stage of the Semantic Genesis framework, where an auto-encoder is trained to reconstruct the input images. Such steps help in discovering a set of semantically similar patients who share similar anatomical patterns by comparing their encoding vectors. Consequently, a random number of crops with fixed coordinates are derived from those patients and assigned a numerical label that denotes their positions. For the self-classification stage,  a classification head, on the top of the framework encoder, is employed to classify the extracted batches according to their assigned labels. In addition, the same intuition of Models genesis is adopted in the self-restoration phase but applied to the extracted patches rather than the whole image. This way, Semantic Genesis enables learning semantically rich representations from similar anatomical patterns. Seven downstream tasks were utilized for evaluation as classification and segmentation tasks.

\cite{zhang2020universal} introduced scale-aware restoration pretext task for 3D medical images segmentation as an extension of Models Genesis framework \citep{zhou2019models}. In addition to the transformation restoration as in Models Genesis, the authors introduced scale discrimination property to the original model depending on the fact that desired objects, e.g. tumors, appear in different sizes across different patients. And hence, cubes of predefined sizes as small, medium, and large are generated and resized into a unified size and then labeled according to their original cropping size. Consequently, the classification head is included on the top of the encoder to accomplish the scale classification task; while the whole architecture is responsible for the transformation restoration task. Brain tumor segmentation and pancreas organs and tumors segmentation were used as downstream tasks. 

\cite{dong2021self} developed a multi-task self-supervised learning approach that combines both generative modeling and instance discrimination using sequential medical data. Given a sequence of medical images for the same patient e.g CT, an auto-encoder architecture, with a single encoder and two decoders, is responsible for learning representations by predicting the T steps precedent and successor slices of the input slice which in turn enables learning the anatomical structural similarity between different slices. In addition, an instance discrimination task is included to avoid learning trivial features by generative modeling. To achieve this, an additional encoder is introduced to the whole architecture that takes another input slice from the same patient and tries to contrast it with the generative model input by minimizing the negative cosine similarity between both inputs. It is worth noting that the second encoder shares the same weights with the generative model whereas it does not go through the back-propagation process.

\cite{koohbanani2021self} proposed a self-path framework for histopathology images which comprises three pathology-specific tasks, namely, magnification prediction, magnification Jigsaw puzzle (JigMag), and Hematoxylin channel prediction in multi-tasking settings. For the first one, patches with different predefined levels of magnification are extracted whereas the task role is to predict the right magnification of the input image. For JigMag, the generated puzzles for training include patches with different magnifications levels for the same image, while the task role is to predict the right order of the puzzle. For the latter task, out of a histopathological image stained with Hematoxylin and Eosin, the role of the task is to predict the first channel from the stained image. Lastly, all proposed tasks along with the downstream tasks are trained jointly in a multi-tasking fashion.

\cite{zhang2021twin} developed a semi-supervised multi-tasking approach that combines rotation prediction \citep{komodakis2018unsupervised}, Jigsaw puzzle \citep{noroozi2016unsupervised}  and SimCLR \citep{chen2020simple} in a unified framework called twin self-supervision based semi-supervised learning (TS-SSL) for spectral-domain optical coherence tomography (SD-OCT) classification. For the Jigsaw puzzle, the authors introduced patch rotation as given in \citep{li2020self}, while for SimCLR the authors introduced supervised category-wise contrastive loss, which considers all samples for a certain label as positive examples. Consequently, the proposed approach is trained in an end-to-end fashion and semi-supervised multitasking setting to learn representations by performing rotation prediction, Jigsaw puzzle-solving, contrastive and supervised contrastive learning. The methods were evaluated on multi-class and binary OCT classification tasks.

\cite{li2021rotation} suggested that rotation-oriented collaborative features learning would provide a potent representation for fundus disorders. They simultaneously combined rotation prediction \citep{komodakis2018unsupervised} with multi-view instance discrimination \citep{wu2018unsupervised} to learn rotation-related and rotation-invariant features using fundus color photography in an end-to-end fashion. Their approach was tested on two ophthalmic diseases, namely, pathological myopia (PM) and age-related macular degeneration (AMD) as a binary classification downstream task. Further, their approach showed that the collaborative approach provided better results than using a single pretext task at a time.

\cite{lu2021volumetric} designed two domain-specific pretext tasks for white matter tract segmentation from diffusion MRI scans. The first task is concerned with predicting the fiber streamlines density map of the white matter in the human brain which represents the number of streamlines that pass through a voxel. On the other side, the second task is concerned with imitating registration-based white matter tract segmentation by registering the input data to a predefined white matter tract registration atlas. Further, both tasks are employed sequentially rather than independently as each of the proposed methods focuses on part of the white matter properties, and hence, integrating them may provide complementary information.  

Table \ref{table6} summarizes the multiple-tasks/multi-tasking self-supervised learning methods in medical imaging.

\begin{table}[!ht]
	\centering
	\begin{tabular}{l|l|l|l}
		No.& Authors     & Pretext task & Down-stream task \\
		\hline

	    1&\cite{tajbakhsh2019surrogate}  &  \vtop{\hbox{\strut Colorization}\hbox{\strut Rotation prediction}\hbox{\strut 3D patch reconstruction}} &
		   \vtop{\hbox{\strut Lung lobe segmentation}\hbox{\strut FPR for nodule detection}\hbox{\strut Skin lesions segmentation}\hbox{\strut Diabetic retinopathy grading}} \\

		 2&\cite{jiao2020self} & \vtop{\hbox{\strut Temporal order correction }\hbox{\strut Transformation prediction}}  &  \vtop{\hbox{\strut Standard plane detection }\hbox{\strut Saliency Prediction}}   \\

		3&\cite{li2020multi} & ColorMe &  \vtop{\hbox{\strut  Cervix type classification }\hbox{\strut Skin lesion segmentation}} \\

		4&\cite{NEURIPS2020_d2dc6368} &\vtop{\hbox{\strut CPC}\hbox{\strut Jigsaw puzzle}\hbox{\strut Exemplar CNN}\hbox{\strut Rotation Prediction}\hbox{\strut Relative position prediction}}  &
		   \vtop{\hbox{\strut Brain tumors segmentation}\hbox{\strut Pancreas tumor segmentation}}  \\

		 5&\cite{luo2020retinal} & Self-supervised fuzzy clustering &  \vtop{\hbox{\strut Color fundus classification}\hbox{\strut  Diabetic retinopathy classification}} \\

		 6&\cite{haghighi2020learning} & Semantic Genesis &  \vtop{\hbox{\strut Lung nodule segmentation}\hbox{\strut FPR for nodule detection}\hbox{\strut Liver segmentation}\hbox{\strut Chest diseases classification}
		     \hbox{\strut Brain tumor segmentation}\hbox{\strut Pneumothorax segmentation}} \\

		 7&\cite{zhang2020universal} & Scale-aware restoration  &  \vtop{\hbox{\strut Brain tumor segmentation}\hbox{\strut Pancreas segmentation}}\\

        8&\cite{dong2021self} & Multi-task self-supervised learning &  Whole heart segmentation  \\

	    9&\cite{koohbanani2021self} & Self-path &  histopathology image classification  \\

		10&\cite{zhang2021twin} & \vtop{\hbox{\strut  SimCLR  }\hbox{\strut Jigsaw puzzle }} & \vtop{\hbox{\strut  Binary OCT classification }\hbox{\strut  Multi-class OCT classification}} \\

        11&\cite{li2021rotation} & \vtop{\hbox{\strut Rotation prediction}\hbox{\strut  multi-view instance discriminate}} & \vtop{\hbox{\strut PM classification}\hbox{\strut  AMD classification}}  \\

		12&\cite{lu2021volumetric} & \vtop{\hbox{\strut Fiber streamlines density map prediction}\hbox{\strut Registration imitation }} & White matter tract segmentation \\

	\end{tabular}
	\caption{Summary of multiple-tasks/multi-tasking self-supervised learning methods in medical imaging.}
	\label{table6}
\end{table}

\section*{Performance comparison}
This section compares the performance of the proposed self-supervised learning approaches that have been discussed in the previous section. Mainly, the emphasis, in this section, is on two tasks which are images classification and semantic segmentation as these two tasks are the most common tasks in the discussed works. Further, this section reports the performance of the proposed self-supervised learning approaches in medical images in comparison to random initialization and transfer learning from ImageNet where applicable. Lastly, this section considers only the benchmarks that have more than two works evaluated on them.

\subsubsection*{Classification tasks performance comparison}
For the sake of performance comparison of the proposed methods from the previous section on the classification tasks, three datasets are selected in which two of them are public datasets which are Lung Nodule Analysis LUNA \citep{DBLP:journals/corr/SetioTBBBC0DFGG16}, and CheXpert dataset \citep{irvin2019chexpert}, while the third one is a private dataset which is Brain Hemorrhage classification dataset. For consistent comparison, the most common performance measure among all proposed approaches is reported. The area under curve (AUC) for receiver operating characteristic curve (ROC) is reported for both LUNA and CheXpert datasets, while overall accuracy is reported for the brain hemorrhage classification dataset. Further, all reported results are in terms of fine-tuning the whole model on the downstream task rather than the results of the fine-tuning the classification layer.
Table \ref{table7} reports the results on the LUNA dataset while Table \ref{table8} reports the results on the brain hemorrhage classification dataset, and Table \ref{table9} reports the results on the CheXpert dataset.

It can be clearly observed from tables \ref{table7}, \ref{table8} and \ref{table9} that self-supervised learning approaches provide better performance when compared to either training from scratch or using ImageNet pre-trained models. In addition, designing new approaches or modifying existing methods in a way that considers the properties of medical images provides better performance than directly adopting pretext tasks from the computer vision field. Contrastive learning based algorithms tend to show better results when compared to other categories as shown in Table \ref{fig7}. However, this is not always the case where the TCPC approach showed the best and the worst results on the brain hemorrhage classification dataset when using different backbones as shown in Table \ref{table8}. In Table \ref{table9}, it can be clearly seen that directly adopting pretext task from computer vision to the medical images analysis field provides marginal improvements on the down-stream task as compared to the ImageNet pre-trained models as given by both  \cite{pmlr-v143-sowrirajan21a} and  \cite{azizi2021big}. On the other side, modifying computer vision tasks by incorporating medical knowledge, as given by \cite{vu2021medaug}, provided significant improvements on the performance when compared to ImageNet pre-trained models.

\begin{table}[!ht]
	\centering
	\begin{tabular}{l|l|l|l|l|l}
		No.& Author     & Pretext task & Category & Random init.: AUC & SSL: AUC \\
		\hline
		
		1  &  \cite{zhu2020embedding}  & TCPC$^{**}$  & Contrastive & 0.982 & 0.996 \\	
		
		2  &  \cite{zhu2020embedding}  & TCPC$^{**}$  & Contrastive & 0.911 & 0.987 \\	
		
		3  &  \cite{haghighi2020learning}  &  Semantic Genesis & Multi-tasking & 0.943 & 0.985 \\
		
		4  &   \cite{zhou2019models}  & Models Genesis & Generative & 0.942 & 0.982 \\

		5  &  \cite{haghighi2020learning}  & Rubik Cube$^*$  & Predictive &  0.943  & 0.955 \\

		6  &  \cite{haghighi2020learning}  &  Context Restoration$^*$  & Generative &  0.943  & 0.919 \\

		7  &  \cite{haghighi2020learning}  &  Image Inpainting$^*$  &  Generative &  0.943  & 0.915 \\
		
		8  &  \cite{haghighi2020learning}  &  Auto-encoder$^*$ & Generative &  0.943  & 0.884 \\

		9  &  \cite{tajbakhsh2019surrogate}  &  3D patch Reconstruct. & Generative & 0.724 & 0.739 \\	
		
	\end{tabular}
	\caption{Performance comparison on LUNA 2016 dataset. Pretext tasks indicated with * are reproduced by the same author. Pretext tasks indicated with ** are implemented using different backbones.}
	\label{table7}
\end{table}

\begin{table}[!ht]
	\centering
	\begin{tabular}{l|l|l|l|l|l}
		No.& Author     & Pretext task & Category & Random init.: Acc (\%)  & SSL: Acc (\%) \\
        \hline
		
		1  & \cite{zhu2020embedding} & TCPC$^{**}$ & Contrastive & 81.08 & 88.17 \\
		
		2  &  \cite{zhu2020rubik}  & Rubik Cube+$^{**}$ & Predictive & 79.73 & 87.84 \\	
		
		3  &  \cite{zhuang2019self}  & Rubik Cube & Predictive &  72.60 &  83.80\\	
		
		4  &  \cite{zhu2020rubik}  & Rubik Cube+$^{**}$ & Predictive & 72.30 & 78.68 \\	
		
		5  & \cite{zhu2020embedding} & TCPC$^{**}$ & Contrastive & 72.30 & 78.38 \\

	\end{tabular}
	\caption{Performance comparison on brain hemorrhage classification dataset. Pretext tasks indicated with ** are implemented using different backbones.}
	\label{table8}
\end{table}

\begin{table}[!ht]
	\centering
	\begin{tabular}{l|l|l|l|l|l}
		No.& Author     & Pretext task & Category & ImageNet: AUC  & SSL: AUC \\
		\hline
		
		1  &  \cite{pmlr-v143-sowrirajan21a}  & MoCo  & Contrastive & 0.949 & 0.953 \\	
		
		2  &  \cite{vu2021medaug}  & MoCo  & Contrastive & 0.858  & 0.906 \\	
		
		3  &  \cite{azizi2021big}  & SimCLR & Contrastive & 0.763 & 0.767 \\

	\end{tabular}
	\caption{Performance comparison on CheXpert dataset.}
	\label{table9}
\end{table}

\subsubsection*{Segmentation tasks performance comparison}
Performance comparison on semantic segmentation tasks is reported on brain tumor segmentation dataset (BraTS 2018) \citep{bakas2018identifying}. All listed works are compared with respect to the Dice Score (DSC) as it is the most common performance among all works on the same dataset. Lastly, all reported results are in terms of fine-tuning the whole model on the downstream task.

Table \ref{table10} reports the results of the BraTS dataset. It can be clearly seen that the performance of self-supervised learning based approaches significantly outperforms training from scratch. In addition, it can be observed that the adopted methods from the computer vision field do not provide better performance as compared to those methods designed especially to suit the nature of medical images.

\begin{table}[!ht]
	\centering
	\begin{tabular}{l|l|l|l|l|l}
		No.& Author & Pretext task & Category & Random init.: DSC (\%) & SSL: DSC (\%) \\
		\hline
		1  &  \cite{zhou2019models}  & Models Genesis & Generative & 90.68 & 92.58 \\
		
		2  &  \cite{taleb2021multimodal}  & Jigsaw Puzzle & Predictive & 80.54 & 89.74\\	
		
		3  &  \cite{zhu2020rubik}  & Rubik Cube+ & Predictive & 85.47 & 89.6 \\	
		
		4  & \cite{chen2019self} & Context Restoration & Generative & 84.41 & 85.57 \\
		
		5  & \cite{zhang2020universal} & Scale Aware Rest. & Multi-Tasking & 74.35 & 84.92 \\
		
		6  & \cite{chen2019self} & Image Inpainting$^*$  & Generative & 84.41 & 84.54 \\		
		
		7  & \cite{NEURIPS2020_d2dc6368} & 3D Relative Pos. Pred. & Predictive & 76.38 & 81.28 \\
		
		8  & \cite{NEURIPS2020_d2dc6368} & 3D CPC & Contrastive & 76.38 & 80.83 \\
		
		9  & \cite{NEURIPS2020_d2dc6368} & 3D Rotation  & Predictive & 76.38 & 80.21 \\
		
		10  & \cite{NEURIPS2020_d2dc6368} & 3D Jigsaw & Predictive & 76.38 & 79.66 \\
		
		11  & \cite{NEURIPS2020_d2dc6368} & 3D Exemplar & Predictive & 76.38 & 79.46 \\

	\end{tabular}
	\caption{Performance comparison based on BraTS dataset.}
	\label{table10}
\end{table}

\section*{Discussion and future research directions}
\label{sec:disc}

A variety of methods that employed self-supervised learning in medical imaging analysis have been discussed in the previous section with respect to each category. Some of these works adopted methods from the computer vision field.
Other researchers proposed novel approaches by incorporating medical knowledge into the design of the pretext task or exploiting the unique properties of the medical images. This section discusses the salient insights that can be derived from the previously discussed works.

\textbf{Computer vision task in medical imaging:} Despite the fact that computer vision and medical images analysis fields deal with image data, there are fundamental differences in the characteristics of natural images and medical images in terms of the number of channels, intensity, location, scale, and orientation. For the number of channels, natural images are mainly 2D RGB images, while medical images may be 2D gray-scale, 3D volumes, or 4D as volume over time dimension. For intensity, the same object will nearly possess the same features under different intensity levels, e.g., the human face is the same under different intensities. On the other side, intensity holds meaningful information in medical images, e.g., different tissues have different values according to the Hounsfield scale in CT scans. For the location, objects in natural images are not affected by changing locations, e.g., the human face holds the same features for the same person in different locations. In medical images, object location has significant indication with respect to certain pathology, e.g., Diabetic Macular Edema severity is diagnosed by examining the Oedema presence with respect to the Fovea in OCT scans. For scale, an object's features in natural images are not significantly affected by the scale, e.g., the human face will not change significantly by changing magnification levels. In contrast, scale is an important factor in some medical imaging modalities, e.g., in histopathological images, different information can be obtained at different magnification levels. Eventually, orientation in natural images is a significant factor for some applications, e.g., texts and numbers orientation in optical character recognition applications. In medical images,  orientation may not be a decisive factor, e.g., tumors may have different non-predefined shapes which in turn make them agnostic for orientations. In summary, medical images have unique properties that distinguish them from the natural images that need to be considered \citep{zhou2021review}.

The direct adoption of pretext tasks from the computer vision field, which have achieved state-of-the-art results on natural images, may not necessarily give the same performance in the medical images analysis field. Hence, knowing that medical images have unique properties as compared to natural images, these properties must be taken into consideration when adopting pretext tasks from computer vision. 

A variety of the discussed works considered the unique properties of medical images when adopting pretext tasks from the computer vision field and modified these methods accordingly. 
For instance, several works modified the existing methods to be able to deal with the volumetric nature of the medical images rather than 2D images such as in \citep{NEURIPS2020_d2dc6368,zhuang2019self,zhu2020rubik, zhu2020embedding}. Other researchers provided modifications to the existing methods to suit the nature of medical images in terms of loss functions such as in   \citep{NEURIPS2020_949686ec,li2020self-sup,xie2020pgl}, and positive pairs selection for contrastive learning algorithms such as in \citep{jamaludin2017self,azizi2021big,NEURIPS2020_949686ec,vu2021medaug,li2020self-sup}. Lastly, some researchers combined more than one computer vision task together to enable robust representations learning such as in \citep{zhang2021twin,li2021rotation}. To sum up, pretext tasks adopted from computer vision need to be modified when adopted in medical imaging analysis in a way that fits the unique characteristics of medical images. Further, the previously mentioned differences between natural images and medical images can be considered as design considerations for the research in the field.

\textbf{Pretext tasks based on medical knowledge:}
Most of the presented works that proposed novel pretext tasks tend to be based on the manipulation of the input image as well as the property of the images. Fewer works tend to incorporate medical knowledge into their approaches such as in \citep{hu2020self,lu2021volumetric, hervella2020multi,holmberg2020self,vu2021medaug}. This may be attributed to the fact that incorporating medical knowledge such as patient metadata, cross-modal images, and disease-specific knowledge may limit the applicability of the proposed self-supervised learning approach to a certain imaging modality and specific disease and may limit its transferability to other tasks without the need to modify the core of the proposed approach. On the other hand, exploiting medical images' properties as well as images manipulation as the bases for the design of pretext tasks provides a wider range of applications for different imaging modalities that possess common attributes. Medical knowledge incorporation with the design of pretext tasks is another research direction that needs to be further explored to benefit from such available knowledge in designing self-supervised learning approaches that are able to provide robust representations empowered by the medical knowledge.

\textbf{Pretext tasks design with multiple imaging modalities:} Diagnosis of a certain disease or capturing a certain organ in the clinical practice may be performed by using more than one imaging modality as they provide complementary information. As an example, OCT scans and fundus color photos are used to diagnose retina diseases. Several works that have been discussed in this survey considered such property in their designs of pretext tasks such as in \citep{holmberg2020self,hervella2020learning,li2020self-sup, taleb2021multimodal}. While learning from a single imaging modality can produce good representations, incorporating multiple imaging modalities, in the design of pretext tasks, can offer learning rich representations. Hence, additional research efforts need to be performed in this direction.

\textbf{Data availability:} Most of the presented works utilize either public or private datasets for the training phase of the pretext tasks. Public datasets are known to be of small size except for some modalities such as X-ray \citep{irvin2019chexpert,wang2017chestx}, fundus\footnote{https://www.kaggle.com/c/diabetic-retinopathy-detection} color photo, and optical coherence tomography \citep{kermany2018large} which are available with a considerable number of images. On the other side, private datasets are not available to the research community and are not easy to reach. Hence, there is a need for building large unlabeled data pools that cover a wide range of imaging modalities to be available for the research community to accelerate the application of self-supervised learning in the field. An important point to consider, when developing an unlabeled medical images dataset, is the data bias. More clearly, medical images datasets in general and medical images datasets in specific tend to be biased toward the healthy cases, while fewer images represent the abnormalities. Data bias must be avoided when developing self-supervised learning methods to guarantee learning representations that are rich in pathological features.

\section*{Conclusion}
\label{sec:conc}

Machine learning applications in medical imaging analysis require large amounts of high-quality annotated data to develop robust models in a supervised fashion, which may not be always available at our disposal. Annotated medical images are scarce and this acts as a major problem that researchers, in the field of machine learning, encounter. Self-supervised learning methods can significantly alleviate the problem of scarcity in annotated data, in the field of medical images analysis, as it enables learning robust representations from unlabeled data.

This is the first survey, to the best of our knowledge, that covers recent self-supervised learning methods and their applications in the field of medical imaging analysis and cast them into four categories, namely, predictive, generative, contrastive, and multi-tasking. This survey extensively reviews 15 state-of-the-art self-supervised learning methods from the computer vision field that have been extensively employed in the context of medical imaging analysis. In addition, the survey covers the 40 most prominent self-supervised learning applications in the field of medical imaging analysis for different imaging modalities and medical conditions. Further, a comparative analysis is conducted to highlight the best performers among the reviewed self-supervised learning approaches in the medical images field when compared on a unified benchmark. Finally, this survey summarizes the major patterns that can be observed from the discussed self-supervised learning applications in medical imaging. Moreover, this survey emphasizes some of the open issues in the field that requires attention from the research community.

\section*{Declaration of competing interests}

The authors have no conflicts of interests to declare that are relevant to the content of this article.

\section*{Appendex A}
\label{appendix:app1}

\setcounter{table}{0}
\renewcommand{\thetable}{A\arabic{table}}
Table \ref{table25} lists the implementation of the previously listed works from both computer vision and medical image analysis who render their code publicly available. Further, starred implementations represents the authors' official code. 

\begin{table}[!ht]
	\centering
	\begin{tabular}{l|l|l|l}
		No.&Authors     & Pretext task &  Implementation \\
		\hline
		
		1&\cite{dosovitskiy2015discriminative}& Exemplar CNN & \href{https://github.com/yihui-he/Exemplar-CNN}{caffe$^*$} \\

		2&\cite{doersch2015unsupervised} &  Relative position prediction & \href{https://github.com/cdoersch/deepcontext}{caffe$^*$} \ \ \ \ |\ \  \href{https://github.com/abhisheksambyal/Self-supervised-learning-by-context-prediction}{pytorch} \\

		3&\cite{noroozi2016unsupervised} & Jigsaw puzzle & \href{https://github.com/bbrattoli/JigsawPuzzlePytorch}{pytorch} \\

		4&\cite{komodakis2018unsupervised} & Rotation prediction & \href{https://github.com/gidariss/FeatureLearningRotNet}{pytorch$^*$}  \\

		5&\cite{vincent2008extracting} & Denoising auto-encoder& \href{https://github.com/GuruMulay/stacked-denoising-autoencoder}{theano} \ \ \ |\ \  \href{https://github.com/shadow2496/KAIST_2019_Deep-Learning_HW3}{pytorch} \\

		6&\cite{pathak2016context} & Image inpainting & \href{https://github.com/pathak22/context-encoder}{caffe$^*$} \ \ \ \ |\ \  \href{https://github.com/jazzsaxmafia/Inpainting}{tensorflow} \\

		7&\cite{zhang2016colorful} & Image colorization & \href{https://github.com/richzhang/colorization}{pytorch$^*$}  \\

		8&\cite{zhang2017split} & Split-brain auto-encoder & \href{https://github.com/richzhang/splitbrainauto}{caffe$^*$} \ \ \ \ |\ \  \href{https://github.com/ysharma1126/Split-Brain-Autoencoder}{tensorflow}  \\

		9&\cite{DBLP:unsupjournals/corr/RadfordMC15} & Deep Convolutional GAN & \href{https://github.com/eriklindernoren/PyTorch-GAN#deep-convolutional-gan}{pytorch} \ \ |\ \  \href{https://github.com/carpedm20/DCGAN-tensorflow}{tensorflow} \\

		10&\cite{donahue2016adversarial} & Bi-directional GAN & \href{https://github.com/jeffdonahue/bigan}{theano$^*$} \ \ |\ \  \href{https://github.com/eriklindernoren/Keras-GAN#bigan}{tensorflow} \\

		11&\cite{DBLP:cpcjournals/corr/abs-1807-03748} & CPC & \href{https://github.com/jefflai108/Contrastive-Predictive-Coding-PyTorch}{pytorch} \ \ \ \ \ \ \ \ |\ \  \href{https://github.com/davidtellez/contrastive-predictive-coding}{tensorflow} \\

		12&\cite{he2020momentum} & MoCo  &\href{https://github.com/facebookresearch/moco}{pytorch$^*$} \ \ \ \ \ \ |\ \  \href{https://github.com/ppwwyyxx/moco.tensorflow}{tensorflow$^*$}  \\

		 13&\cite{chen2020simple} & SimCLR & \href{https://github.com/google-research/simclr}{tensorflow$^*$} \ |\ \  \href{https://github.com/sthalles/SimCLR}{pytorch}  \\

		14&\cite{NEURIPS2020_f3ada80d} & BYOL & \href{https://github.com/deepmind/deepmind-research/tree/master/byol}{tensorflow$^*$} \ |\ \  \href{https://github.com/lucidrains/byol-pytorch}{pytorch}  \\

		 15&\cite{NEURIPS2020_70feb62b} & SwAV & \href{https://github.com/facebookresearch/swav}{pytorch$^*$} \ \ \ \ \ \ |\ \  \href{https://github.com/ayulockin/SwAV-TF}{tensorflow} \\

		  16&\cite{NEURIPS2020_d2dc6368}&\vtop{\hbox{\strut CPC}\hbox{\strut Jigsaw puzzle}\hbox{\strut Exemplar CNN}\hbox{\strut Rotation Prediction}\hbox{\strut Relative position prediction}} & \href{https://github.com/HealthML/self-supervised-3d-tasks}{tensorflow$^*$} \\

		  17&\cite{li2021rotation} & \vtop{\hbox{\strut Rotation prediction}\hbox{\strut  multi-view instance discriminate}} &  \href{https://github.com/xmengli999/Rotation-oriented-self-supervised}{tensorflow$^*$} \\

		  18&\cite{sriram2021covid} & MoCo & \href{https://github.com/stanfordmlgroup/MoCo-CXR}{pytorch$^*$}  \\

		  19&\cite{pmlr-v143-sowrirajan21a} & MoCo & \href{https://github.com/stanfordmlgroup/MoCo-CXR}{pytorch$^*$}  \\

		  20&\cite{xie2020pgl} & BYOL & \href{https://github.com/YtongXie/PGL}{pytorch$^*$} \\

		  21&\cite{NEURIPS2020_949686ec} & SimCLR & \href{https://github.com/krishnabits001/domain\_specific\_cl
            }{tensorflow$^*$}   \\

		  22&\cite{zhou2019models} & Models Genesis & \href{https://github.com/MrGiovanni/ModelsGenesis}{tensorflow$^*$} \\

		 23&\cite{haghighi2020learning} & Semantic Genesis & \href{https://github.com/fhaghighi/SemanticGenesis}{tensorflow$^*$} \\

		 24&\cite{li2020self-sup}  & Feature-based softmax embedding & \href{https://github.com/xmengli999/self_supervised}{pytorch$^*$} \\

		   25&\cite{holmberg2020self} & Cross modal retinal thickness prediction & \href{https://github.com/theislab/DeepRT}{tensorflow$^*$}  \\

		  26&\cite{matzkin2020self} &  skull reconstruction & \href{https://gitlab.com/matzkin/headctools}{pytorch$^*$} \\

		  27&\cite{zhang2021twin} &  TS-SSL & \href{https://github.com/ZhangYH0502/TS-SSL}{tensorflow$^*$} \\

		   28&\cite{prakash2020leveraging} &  Images denoising & \href{https://github.com/juglab/VoidSeg}{tensorflow$^*$} \\

	\end{tabular}
	\caption{Implementation codes list }
	\label{table25}
\end{table}

\bibliography{sample}

\end{document}